\documentclass[
 reprint,
 superscriptaddress,
 amsmath,
 amssymb,
 aps,
 prl,
 floatfix,
 nofootinbib
]{revtex4-2}

\usepackage{graphicx}
\usepackage{dcolumn}
\usepackage{bm}
\usepackage{subfig}
\usepackage{float}
\usepackage{multirow}
\usepackage{adjustbox}
\usepackage{booktabs, caption, makecell}

\usepackage{threeparttable}
\usepackage{caption}
\usepackage{siunitx}
\usepackage{upgreek}
\usepackage{xcolor}
\usepackage{lipsum}
\usepackage{mathrsfs}
\usepackage{amsfonts}
\usepackage{mathtools}
\usepackage{cancel}

\newcommand{\ha}{\hat{a}}
\newcommand{\hb}{\hat{b}}
\newcommand{\hc}{\hat{c}}
\newcommand{\hd}{\hat{d}}
\newcommand{\hf}{\hat{f}}
\newcommand{\hg}{\hat{g}}

\newcommand{\di}{\mathrm{d}}
\newcommand{\deff}{{\, \vcentcolon = \,}}

\newcommand{\mm}{{m m}}

\newcommand{\nm}{{n m}}

\newcommand{\maR}{\mathcal{R}\,}
\newcommand{\mT}{\mathcal{T}}

\usepackage[normalem]{ulem}

\providecommand{\abs}[1]{\left|#1\right|}

\usepackage{ragged2e} 
\makeatletter
\renewcommand{\@makecaption}[2]{%
  \vskip\abovecaptionskip
  \sbox\@tempboxa{#1: #2}%
  \ifdim \wd\@tempboxa >\hsize
    \justifying #1: #2\par
  \else
    \global \@minipagefalse
    \hb@xt@\hsize{\hfil\box\@tempboxa\hfil}%
  \fi
  \vskip\belowcaptionskip}
\makeatother

\begin{document}

\preprint{APS/123-QED}

\title{Hong--Ou--Mandel-like interference between two independent narrowband thermal states}

\author{Sheng-Hsuan Huang}
 \email{sheng-hsuan.huang@mpl.mpg.de}
\affiliation{Department of Physics, Friedrich-Alexander-Universität Erlangen-Nürnberg, Staudtstrasse 7/A3, 91058 Erlangen, Germany} 
\affiliation{Max Planck Institute for the Science of Light, Staudtstrasse 2, 91058 Erlangen, Germany}
\author{Thomas Dirmeier}
\affiliation{Department of Physics, Friedrich-Alexander-Universität Erlangen-Nürnberg, Staudtstrasse 7/A3, 91058 Erlangen, Germany} 
\affiliation{Max Planck Institute for the Science of Light, Staudtstrasse 2, 91058 Erlangen, Germany}
\author{Andrea Aiello}
\affiliation{Max Planck Institute for the Science of Light, Staudtstrasse 2, 91058 Erlangen, Germany}
\author{Kaisa Laiho}
\affiliation{German Aerospace Center (DLR e.V.), Institute of Quantum Technologies, Wihelm-Runge-Str. 10, 89081 Ulm, Germany}
\author{\\Dmitry V. Strekalov}
\affiliation{Max Planck Institute for the Science of Light, Staudtstrasse 2, 91058 Erlangen, Germany}
\author{Gerd Leuchs}
\affiliation{Department of Physics, Friedrich-Alexander-Universität Erlangen-Nürnberg, Staudtstrasse 7/A3, 91058 Erlangen, Germany} 
\affiliation{Max Planck Institute for the Science of Light, Staudtstrasse 2, 91058 Erlangen, Germany}
\author{Christoph Marquardt}
\affiliation{Department of Physics, Friedrich-Alexander-Universität Erlangen-Nürnberg, Staudtstrasse 7/A3, 91058 Erlangen, Germany}
\affiliation{Max Planck Institute for the Science of Light, Staudtstrasse 2, 91058 Erlangen, Germany} 

\date{\today}

\begin{abstract}

Narrowband photonic sources are promising candidates in quantum networks for a direct interface with atomic quantum memories. However, building a large-scale quantum network requires many indistinguishable sources, and matching the frequencies of independent narrowband sources is generally challenging. Here, we propose a method to measure the frequency difference between two narrowband light sources in the single-photon power regime by studying Hong–-Ou–-Mandel-like (HOM) interference between thermal states. Our results show that the absolute value of the frequency difference can be extracted from the temporal coincidence counts, and these results agree well with our theoretical model. We also demonstrate that the beat signal of the HOM-like interference between two thermal states has a similar oscillation profile to the HOM interference between two heralded single-photon sources, confirming that this method can be used to find the zero detuning of two heralded single-photon sources.
\end{abstract}

\maketitle

\section{\label{sec:level1}Introduction}
Interference of photons plays an important role in many quantum applications, such as entanglement generation~\cite{kwiat1995new,huang2024polarization}, quantum communication~\cite{brendel1999pulsed,inoue2002differential,kao2023field}, and precision metrology~\cite{chen2019hong,Chen26}. In particular, the two-photon interference, known as Hong--Ou--Mandel (HOM) interference~\cite{hong1987measurement}, lies at the heart of many quantum optics tasks. Consequently, it has been widely studied across various photon sources, including single-photon sources~\cite{santori2002indistinguishable, kaltenbaek2006experimental}, weak coherent sources~\cite{rarity2005non, ferreira2015spectral}, and thermal states~\cite{jin2015spectrally}. Indeed, the presence of HOM interference is ubiquitous from entanglement distribution~\cite{pan1998experimental} to quantum computing~\cite{zhong2020quantum, psiquantum2025manufacturable}. Note that much like the Hanbury Brown and Twiss set-up \cite{Loudon}, the Hong-Ou-Mandel set-up can be used to measure intensity interference and correlations for any light field independent of the type of photon statistics.

Meanwhile, narrowband photon sources have drawn a lot of attention due to their capability of being compatible with atomic quantum memories, which are fundamental building blocks for quantum networks. To be able to use narrowband photon sources in large-scale experiments, independent but indistinguishable sources are usually required. However, compared to broadband sources, realizing narrowband indistinguishable sources is generally more challenging due to the stricter frequency-matching requirements. To tackle this challenge, a method for measuring the frequency difference between two photon sources is necessary.

A straightforward approach is to measure the absolute frequencies of the photon sources individually. However, this becomes nontrivial at the single photon level. Conventional frequency measurement techniques, such as wavelength meters or optical spectrum analyzers, often fail due to their limited sensitivity. Further, one may attempt to use spectral filters to estimate the wavelength; however, the achievable spectral precision is typically insufficient for narrowband sources. 

In this work, we present a method for measuring the frequency difference between two narrowband heralded single-photon sources
, implemented through the resonator-assisted spontaneous parametric down-conversion (SPDC). In a weakly-pumped regime, this type of source is known to produce the thermal statistics of both the signal and idler photon populations. Therefore by recording two-fold coincidences we observe the HOM-like interference\footnote{Here, we use 'HOM-like interference' to describe a standard HOM configuration performed with thermal states rather than single photons.} between two thermal states. The time-domain correlation function exhibits a beat signal when a frequency difference is present, and it vanishes as the frequency difference approaches zero. By analyzing the beat signal, we can determine the absolute value of the frequency difference between the two input thermal states. Our experimental results show good agreement with the theoretical model. We further demonstrate a practical use case for two independent heralded single-photons highlighting the advantage of this technique in reducing optimization time. Our results show that careful engineering of the frequency domain of narrowband photonic sources is necessary for embedding them in future quantum networks.

\section{Theory}

\begin{figure}[!ht]
  \centering
  \includegraphics[scale=3,clip=false,width=0.8\columnwidth,trim = 0 0 0 0]{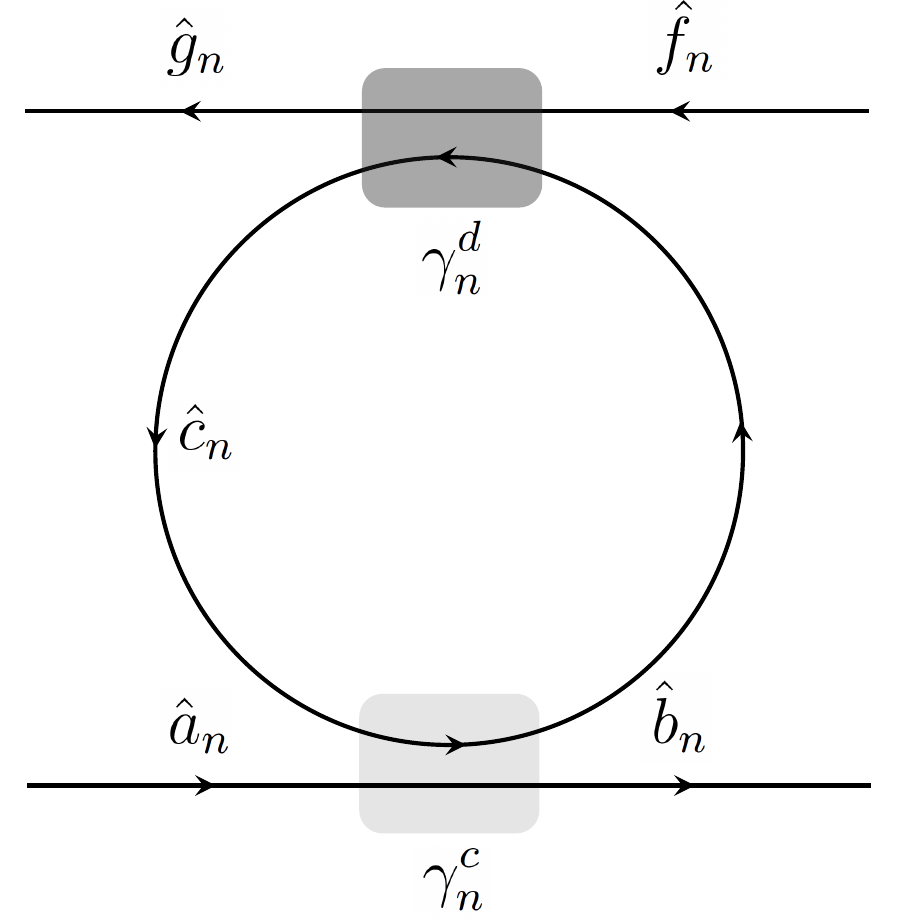}
  \caption{Sketch of the WGMR modeled as a ring cavity. The annihilation operators for the intracavity field $\hc_n(t)$, the input and output external fields $\ha_n(t), \hb_n(t)$, and the input and output noise fields  $\hf_n(t), \,\hg_n(t)$, are described in the text, as are the internal losses rate $\gamma_n^d$, and the outcoupling  rate $\gamma_n^c$, where $n=s,i$.}\label{fig1}
\end{figure}

We produce quantum states of light in a whispering gallery mode resonator (WGMR), in which photon pairs called signal and idler are created via the process of SPDC.
We model our WGMR as a ring cavity, which is sketched in Fig. \ref{fig1}.
\\
%
We consider three electromagnetic waves, denoted as  \emph{pump} ($p$), \emph{signal} ($s$), and  \emph{idler} ($i$). The pump wave is described as a \emph{classical} single-mode field of amplitude $a_p \in \mathbb{C}$, and oscillating at frequency $\omega_p$. The signal and idler waves are represented by multimode \emph{quantum} fields with central frequencies $\omega_s$ and $\omega_i$, respectively. These frequencies are connected by the relation
\begin{align}\label{a20}
\omega_p = \omega_s + \omega_i + \Delta,
\end{align}
where $\Delta$ is the detuning of the pump frequency from the parametric resonance.
The operators and rates depicted in Fig. \ref{fig1} are defined as follows:
\begin{itemize}
  \item $\hc_n(t)$ is the \emph{intra-cavity} photon-number-amplitude operator, such that the expectation value of $\hc_n^\dagger(t)\hc_n(t)$ gives the average number of signal ($n=s$), and idler ($n=i$), photons in the cavity at time $t$.  By construction, $\hc_n(t)$ and  $\hc_n^\dagger(t)$ do not satisfy canonical commutation relations~\cite{ueda1994anomalous}.
  \item $\ha_n(t)$ and $\hb_n(t)$ are the \emph{extra-cavity} input and output field operators, respectively, and $\gamma_n^c$ is the coupling rate between internal and external fields, such that 
\begin{align}\label{a10}
\hb_n(t) = - \ha_n(t) +   \sqrt{\gamma_n^c} \, \hc_n(t), \qquad (n=i, s).
\end{align}
By definition, $\ha_n(t)$ and  $\hb_n(t)$ fulfill the canonical commutation relations. See the Supplementary Material for details.
  \item  $\hf_n(t)$ and $\hg_n(t)$ are the \emph{noise}  input and output field operators, respectively, and $\gamma_n^d$ is the coupling rate between the internal field and the noise fields, such that 
\begin{align}\label{a30}
\hg_n(t) = - \hf_n(t) +   \sqrt{\gamma_n^d} \, \hc_n(t), \qquad (n=i, s).
\end{align}
By definition, also $\hf_n(t)$ and  $\hg_n(t)$ fulfill the canonical commutation relations.
\end{itemize}

Following closely \cite{haus2000electromagnetic} and \cite{shafiee2020nonlinear}, we  write the Heisenberg equations of motion for the  signal and idler modes as, 
\begin{subequations}\label{a70}
\begin{align}
\frac{d \hc_s}{d t}  & = - i \omega_s  \hc_s - i \kappa_1 \, e^{-i \omega_p t}  \hc_i^\dagger - \frac{\Gamma_s}{2} \, \hc_s + \sqrt{\gamma_s^c} \, \ha_s  \nonumber \\[6pt]
& \phantom{=.} - \sqrt{\gamma_s^d} \, \hf_s, \label{a70A} \\[6pt]
\frac{d \hc_i}{d t}  & = - i \omega_i  \hc_i - i \kappa_1 \, e^{-i \omega_p t}  \hc_s^\dagger - \frac{\Gamma_i}{2} \, \hc_i + \sqrt{\gamma_i^c} \, \ha_i \nonumber \\[6pt]
& \phantom{=.} - \sqrt{\gamma_i^d} \, \hf_i, \label{a70B} 
\end{align}
\end{subequations}
where $\Gamma_n \deff \gamma_n^c + \gamma_n^d , \, (n=i, s)$, $\kappa_1 = \kappa \, a_p$, being $\kappa$ the nonlinear coupling rate.

We solve these coupled differential equations using the Fourier transform method \cite{OsgoodBook}. We adopt the following convention (here illustrated for the generic Fourier-pair annihilation operators $\left\{ \ha(t), \, \ha(\omega)\right\}$):
\begin{subequations}\label{a120}
\begin{align}
\ha(t) & = \frac{1}{\sqrt{2 \pi}} \int_\mathbb{R} \ha(\omega) e^{- i \omega t} \di \omega, \label{a120A} \\[6pt]
\ha(\omega) & = \frac{1}{\sqrt{2 \pi}} \int_\mathbb{R} \ha(t) e^{ i \omega t} \di t, \label{a120B} 
\end{align}
\end{subequations}
which implies 
\begin{align}\label{a130}
e^{-i \omega_p t}  \ha^\dagger(t)  = \frac{1}{\sqrt{2 \pi}} \int_\mathbb{R} \ha^\dagger (\omega_p - \omega) e^{- i \omega  t} \di \omega.
\end{align}
Substituting the solutions of \eqref{a70} into \eqref{a10}, we obtain
\begin{subequations}\label{a160}
\begin{align}
\hb_s(\omega) & = A_{ss}(\omega)  \ha_s(\omega)   +  A_{si}(\omega)  \ha^\dagger_i(\omega_p - \omega) \nonumber \\[8pt]
& \phantom{=.} + F_{ss}(\omega)  \hf_s(\omega)  + F_{si}(\omega)   \hf^\dagger_i(\omega_p - \omega)  , \label{a160A} \\[10pt]
\hb_i(\omega) & = A_{ii}(\omega) \ha_i(\omega) + A_{is}(\omega) \ha^\dagger_s(\omega_p - \omega) \nonumber \\[8pt]
& \phantom{=.} +F_{ii}(\omega)  \hf_i(\omega) + F_{is}(\omega) \hf^\dagger_s(\omega_p - \omega)  ,  \label{a160B} 
\end{align}
\end{subequations}
where we have defined
\begin{widetext}
\begin{subequations}\label{a150}
\begin{align}
A_{ss}(\omega) & = \frac{\bigl[\Gamma_i/2  - i \left(\omega - \omega_s - \Delta \right)\bigr] \bigl[\gamma^c_s - \Gamma_s/2  + i \left(\omega - \omega_s \right)\bigr] + \abs{\kappa_1}^2}{\bigl[\Gamma_i/2  - i \left(\omega - \omega_s - \Delta \right)\bigr] \bigl[\Gamma_s/2  - i \left(\omega - \omega_s \right)\bigr] - \abs{\kappa_1}^2}  , \label{a150A} \\[8pt]
A_{si}(\omega)& = \frac{- i \, \kappa_1  \sqrt{\gamma^c_s \gamma^c_i}}{\bigl[\Gamma_i/2  - i \left(\omega - \omega_s - \Delta \right)\bigr] \bigl[\Gamma_s/2  - i \left(\omega - \omega_s \right)\bigr] - \abs{\kappa_1}^2} , \label{a150B} \\[10pt]
F_{ss}(\omega) & = \frac{-\bigl[\Gamma_i/2  - i \left(\omega - \omega_s - \Delta \right)\bigr] \sqrt{\gamma^c_s \gamma^d_s} }{\bigl[\Gamma_i/2  - i \left(\omega - \omega_s - \Delta \right)\bigr] \bigl[\Gamma_s/2  - i \left(\omega - \omega_s \right)\bigr] - \abs{\kappa_1}^2} , \label{a150C} \\[8pt]
F_{si}(\omega)&  = \frac{- i \, \kappa_1  \sqrt{\gamma^c_s \gamma^d_i}}{\bigl[\Gamma_i/2  - i \left(\omega - \omega_s - \Delta \right)\bigr] \bigl[\Gamma_s/2  - i \left(\omega - \omega_s \right)\bigr] - \abs{\kappa_1}^2} ,  \label{a150D} 
\end{align}
\end{subequations}
\end{widetext}
and $\{ A_{ii}, A_{is}, F_{ii}, F_{is} \}$ are obtained by exchanging everywhere the indices $s$ and $i$ in the formulas \eqref{a150} for  $\{ A_{ss}, A_{si}, F_{ss}, F_{si} \}$, respectively. 

\subsection{Second-order correlation functions}

Let us consider now the experimental scenario schematically depicted in Fig. \ref{fig2}. Two independent thermal states carried by the signal beams emitted at frequencies $\omega_{s1}$ and $\omega_{s2}$ by the two WGMRs, enter the input ports of a beam splitter (BS). The latter is characterized by the transmission and reflection coefficients $\mT$ and $\maR$, respectively, with the standard properties:
\begin{figure}[!ht]
  \centering
  \includegraphics[scale=3,clip=false,width=0.8\columnwidth,trim = 0 0 0 0]{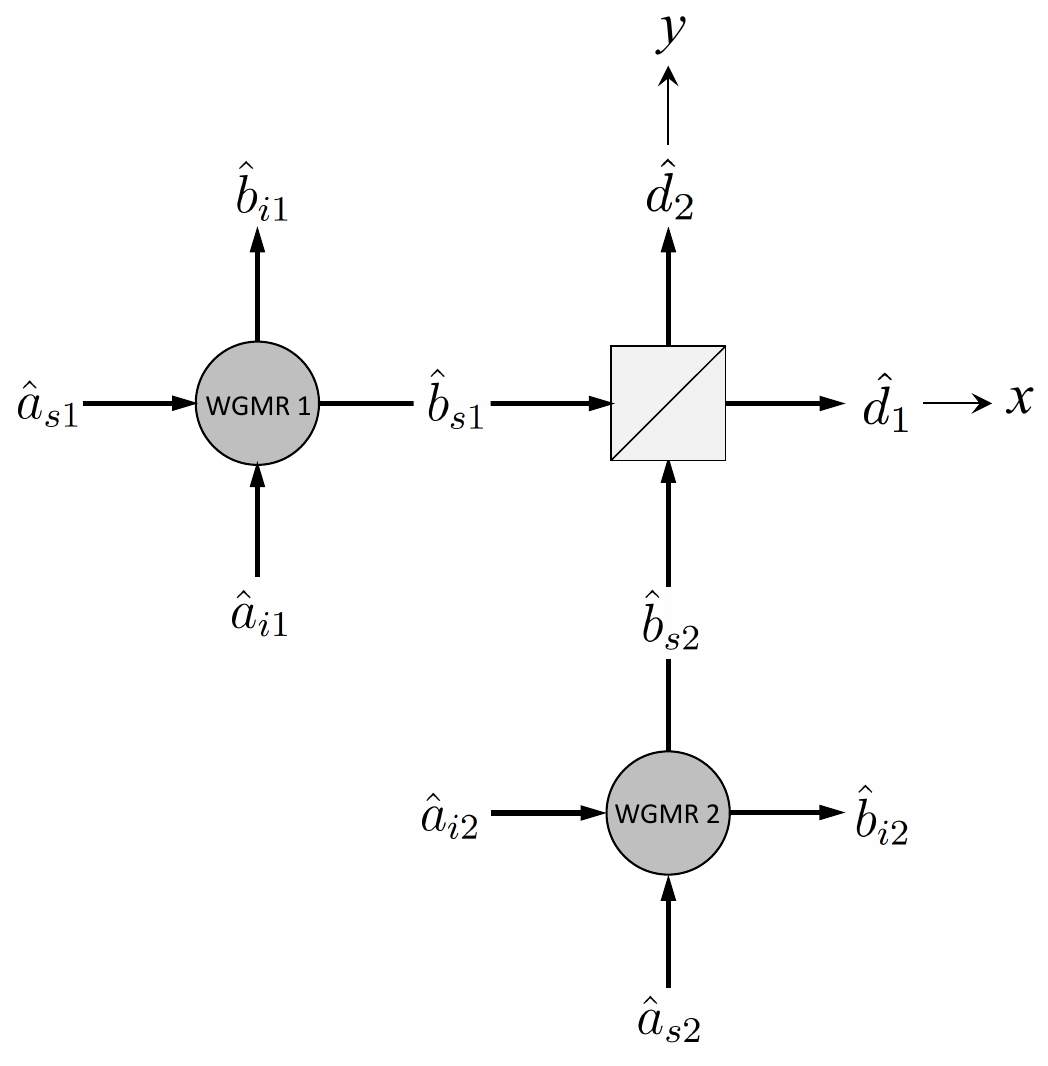}
  \caption{Sketch of the theoretical model of the experimental setup. The ``signal'' beams from two independent WGMRs enters the  ports ``$1$'' and ``$2$'' of a beam splitter (BS), which are  aligned, respectively, to the axes $x$ and $y$ of a Cartesian reference frame. The ``idler'' photons are not detected.}\label{fig2}
\end{figure}
\begin{align}\label{d10}
\abs{\mT}^2 + \abs{\maR}^2 = 1, \qquad \text{and} \qquad \mT \maR^* +  \mT^* \maR = 0.
\end{align}
The BS transforms the annihilation operators $\hb_{s1}(t)$ and $\hb_{s2}(t)$ of the input field into the annihilation operators $\hd_1(t)$ and $\hd_2(t)$ of the output field, according to
\begin{subequations}\label{d20}
\begin{align}
\hd_1(t) & = \mT \, \hb_{s1}(t) + \maR \hb_{s2}(t), \label{d20A} \\[6pt]
\hd_2(t) & = \maR \hb_{s1}(t) + \mT \,  \hb_{s2}(t), \label{d20B}
\end{align}
\end{subequations}
where
\begin{align}\label{d22}
\hb_{s \alpha}(t)  = \frac{1}{\sqrt{2 \pi}} \int_\mathbb{R} \hb_{s \alpha}(\omega) e^{- i \omega t} \di \omega, \quad (\alpha =1,2),
\end{align}
with $\hb_{s \alpha}(\omega)$ being given by \eqref{a160} and \eqref{a150}, where all indexes ``$p$'', ``$i$'' and  ``$s$'' are replaced by ``$p \alpha$'', ``$i \alpha$'' and  ``$s \alpha$'', respectively, and $\Delta \to \Delta_\alpha = \omega_{p\alpha} - \omega_{s\alpha} - \omega_{i \alpha}$, where $\alpha=1,2$ is the beam index.
In our experiment we have zero detuning in both WGMRs, so that $\Delta_1= 0 = \Delta_2$. Moreover, because of the small frequency detuning 
\begin{align}\label{d24}
\delta \omega \deff \omega_{s2} - \omega_{s1},
\end{align}
between the two thermal sources,  the loss and outcoupling rates are  basically frequency-independent, so that $\{\gamma^c_{s \alpha}, \gamma^c_{i\alpha}, \gamma^d_{s\alpha}, \gamma^d_{i\alpha} \}=\{\gamma^c_s, \gamma^c_i, \gamma^d_s, \gamma^d_i \}$.

Two detectors, labelled $D_1$ and $D_2$, are placed after the BS on the axes $x$ and $y$, respectively, at  $x = x_1>0 $ and $y= y_2>0$, where the origin of the coordinates $x=0=y$ is taken at the center of the BS. We  calculate the \emph{interbeam} degree of second-order coherence measured by the detectors $D_1$ and $D_2$, as
\begin{align}\label{d30}
g^{(2)} \left( \tau \right) = \frac{\langle \hd_{1}^\dagger(\tau_1) \hd_{2}^\dagger(\tau_2) \hd_2(\tau_2)\hd_1(\tau_1) \rangle}{\langle \hd_{1}^\dagger(\tau_1) \hd_1(\tau_1) \rangle \langle \hd_{2}^\dagger(\tau_2) \hd_2(\tau_2) \rangle} ,
\end{align}
where  the field operators in the equation above are evaluated  at different times $t_1$ and $t_2$, and we have defined $\tau_1  = t_1 - {x_1}/{c}, \, \tau_2  = t_2 - {y_2}/{c}$, and $\tau = \tau_2 - \tau_1$.
After a lengthy calculation, we obtain
\begin{figure*}[!hbpt]
    \centering
    \includegraphics[width=.8\linewidth]{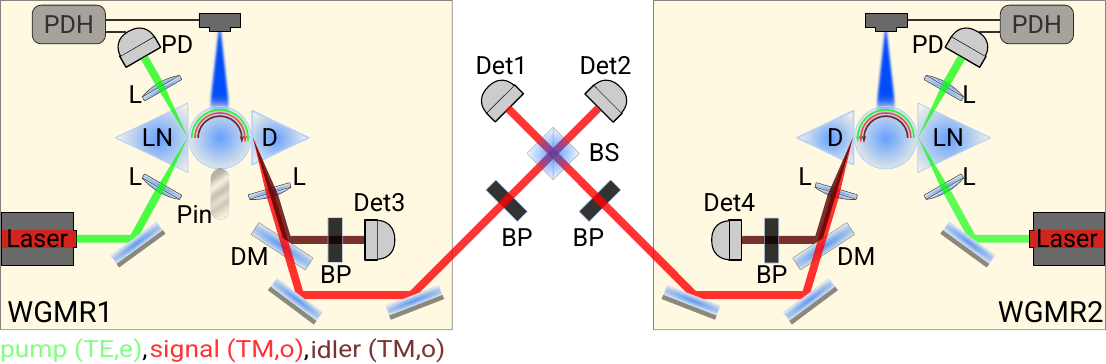}
    \caption{Sketch of the experimental setup. Abbreviations: PDH: Pound–Drever–Hall technique, PD: photodiode, L: lens; LN: x-cut lithium niobate prism, D: diamond prism, DM: dichroic mirror, Pin: aluminum pin, BS: 50:50 beam splitter, BP: bandpass filter, Det1–4: single-photon detectors. TE: transverse electric modes, TM: transverse magnetic modes, o: ordinary, e: extraordinary.}
    \label{fig:setup}
\end{figure*}
\begin{align}\label{d320}
g^{(2)} \left( \tau \right)  = 1 + 2\abs{\maR \mT}^2 \frac{I^2(\tau)}{I^2(0)} \bigl[ 1 - \cos \left(\delta \omega \, \tau \right)\bigr],
\end{align}
where
\begin{align}\label{d330}
I( \tau )  = \frac{1}{2 \pi}\int_\mathbb{R} \frac{e^{- i  x \tau}}{\abs{\left(\Gamma_i/2 - i  x \right) \left(\Gamma_s/2 - i x \right) - \abs{\kappa_1}^2}^2} \di x.
\end{align}
A straightforward calculation gives,
\begin{align}\label{d340}
\frac{I(\tau)}{I(0)} = \frac{a_+ \exp\left(-\dfrac{a_-}{2} \abs{\tau}\right)  - a_- \exp\left(-\dfrac{a_+}{2} \abs{\tau}\right)}{a_+ - a_-},
\end{align}
where
\begin{align}\label{d350}
a_\pm = \sqrt{\frac{\Gamma_{i}^2 + \Gamma_{s }^2 + 8  \kappa_1^2}{2}  \pm \frac{\left( \Gamma_{i} + \Gamma_{s } \right)  
\sqrt{\left( \Gamma_{i} - \Gamma_{s } \right)^2 + 16 \kappa_1^2}}{2} }.
\end{align}
When $\Gamma_i \Gamma_s \gg 4 \kappa_1^2 $, we can set $\kappa_1 =0$ and \eqref{d340} reduces to
\begin{align}\label{d360}
\frac{I(\tau)}{I(0)} = \frac{\Gamma_i \, e^{ - \Gamma_s \abs{\tau}/2 } -  \Gamma_s \, e^{ - \Gamma_i \abs{\tau}/2 } }{\Gamma_i - \Gamma_s}.
\end{align}

\section{Experiment}

A sketch of the experimental setup is shown in Fig.~\ref{fig:setup}. The narrowband thermal states are generated from two WGMRs based on type-I SPDC. Both WGMRs are made of z-cut lithium niobate doped with 5\% MgO, and their shapes are similar to that described in Ref.~\cite{breunig2013whispering}. The first WGMR has a major radius $R \approx \qty{0.75}{\mm}$ and a minor radius $r \approx \qty{0.14}{\mm}$ and is pumped by a \qty{523}{\nm} continuous-wave (cw) laser. The second WGMR has a major radius $R \approx \qty{0.76}{\mm}$ and a minor radius $r \approx \qty{0.13}{\mm}$ and is pumped by a different cw laser at a wavelength of \qty{532}{\nm}. By using two pump lasers we are able to detune the frequency between two thermal states. Both WGMRs are actively temperature-stabilized using Peltier elements in order to achieve desired SPDC emission wavelengths. Two blue lasers are used in each WGMR subsystem as heating lasers for further stabilization~\cite{shafiee2020nonlinear}. The wavelengths of the signals generated from both WGMRs are \qty{935}{\nm}. In addition, an aluminum pin is mounted on top of the first WGMR for fine frequency tuning~\cite{pollinger2009ultrahigh}. See the Supplementary Material for details.

\begin{figure}[!b]
    \centering
    \includegraphics[width=\linewidth]{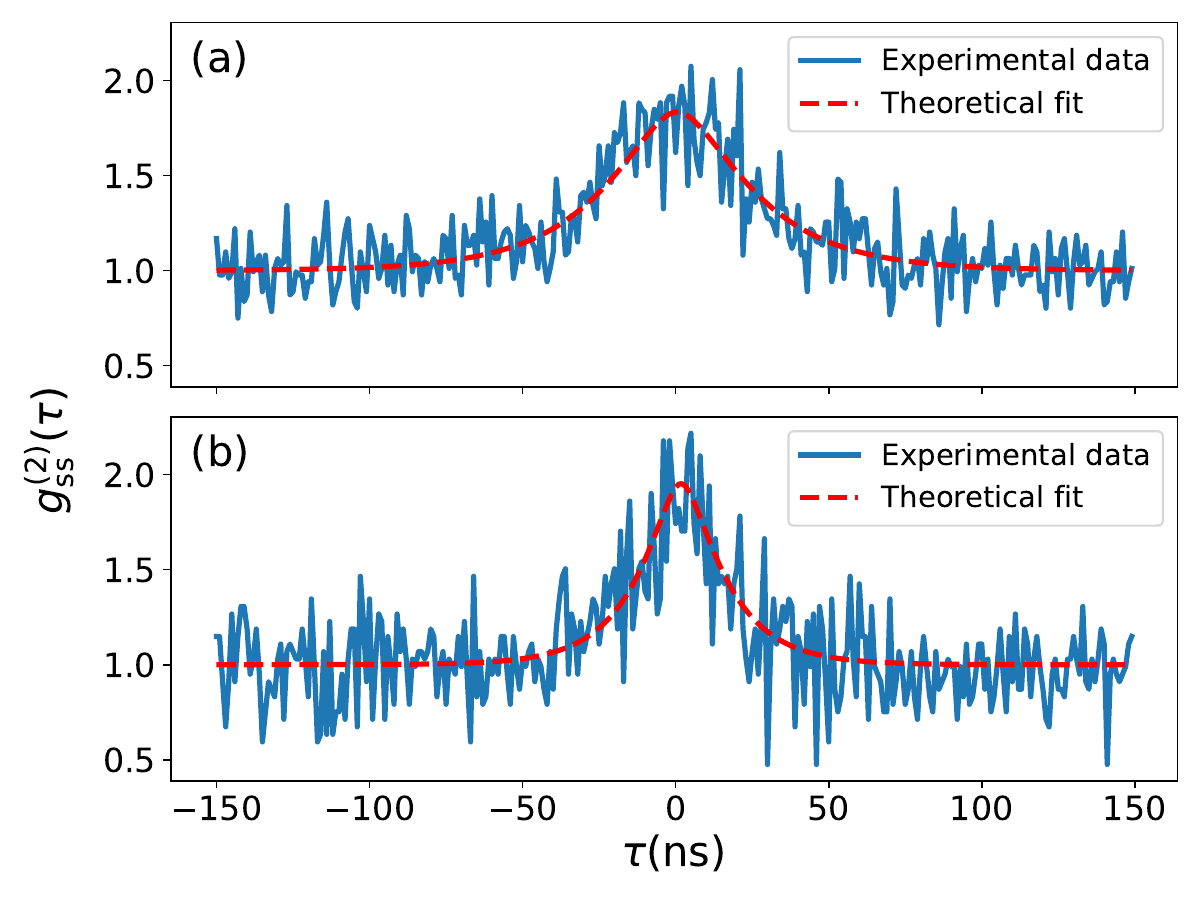}
    \caption{Second-order autocorrelation functions $g^{(2)}_{ss}(\tau)$ measured for the two WGMRs. (a) Data measured from WGMR1. (b) Data measured from WGMR2. The parameter \(\tau\) is the time delay measured between Det1 and Det2. Experimental data are shown in blue, while the red lines present the theoretical fit.}
    \label{fig:auto_g2}
\end{figure}
We first measure the autocorrelation functions, $g^{(2)}_{ss}(\tau)$, of the two independent signal beams to assess the purity of the generated thermal states. The results are shown in Fig.~\ref{fig:auto_g2}. The peak values for the first WGMR is \(1.86\pm0.10\) while that of the second WGMRs is \(1.97\pm0.16\). For a pure single-mode thermal state we expect the peak value of 2~\cite{christ2011probing}. These measurements indicate that the states generated from the WGMRs are close to single-mode thermal states.

\begin{figure}[!ht]
    \centering
    \includegraphics[width=\linewidth]{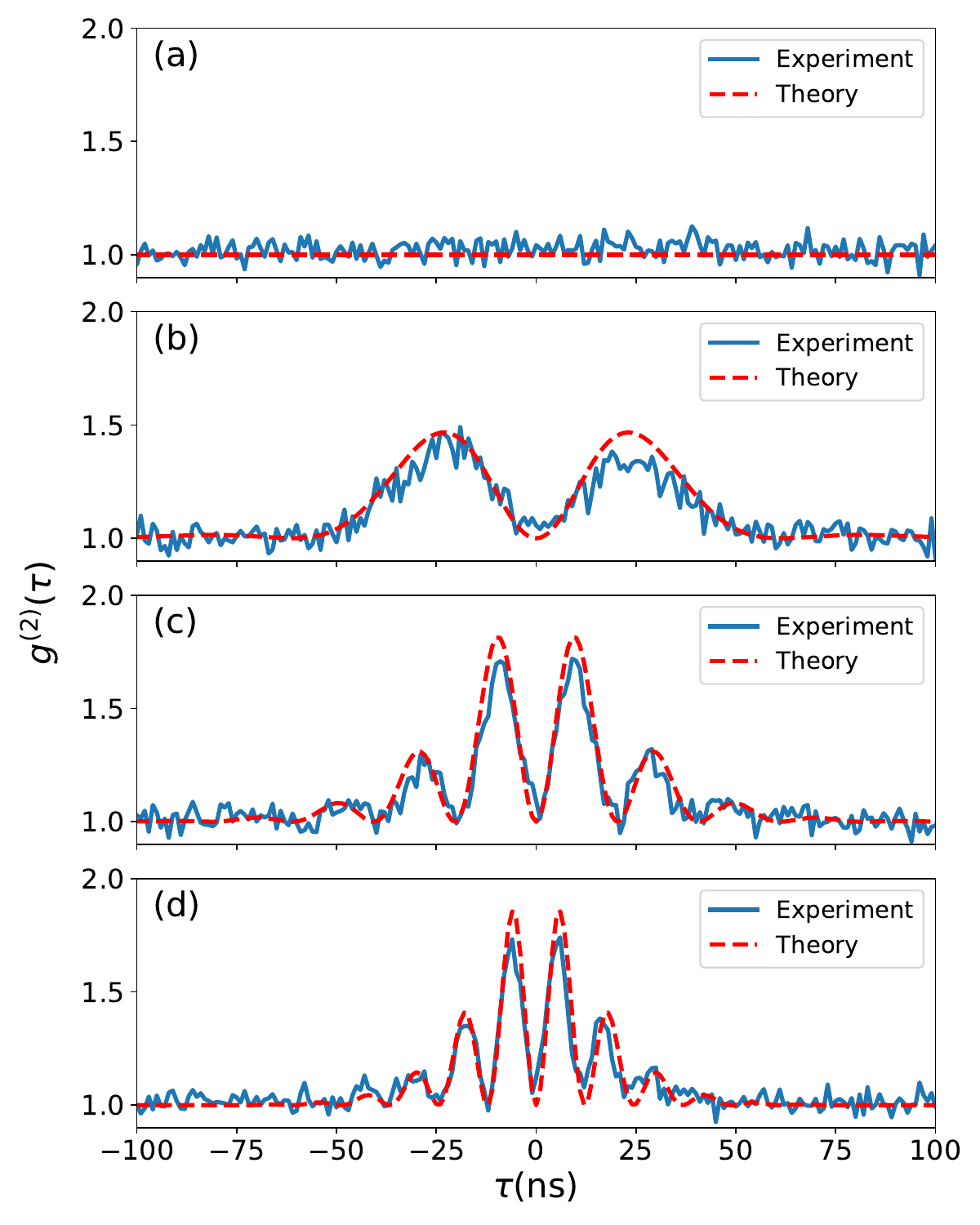}
    \caption{HOM-like interference between two thermal states measured at different frequency detunings as a function of the time delay $\tau$ defined above. (a) Data measured at zero detuning. (b) Data measured at \qty{16}{\MHz} detuning. (c) Data measured at \qty{49}{\MHz} detuning. (d) Data measured at \qty{82}{\MHz} detuning. Experimental data are shown in blue, while the red lines present the theoretical predictions.}
    \label{fig:HOM_t}
\end{figure}

After the thermal states are characterized, the two thermal beams are guided to the two input ports of a symmetric beamsplitter to measure the HOM-like interference. The polarizations of the two thermal states are optimized using polarization controllers, while spatial mode matching is ensured by coupling the light into single-mode fibers. We fix the frequency of the thermal state generated from the second WGMR and vary that of the thermal state generated from the first WGMR and measure four cases with the frequency detuning between two thermal states to be \qty{0}{MHz}, \qty{16}{MHz}, \qty{49}{MHz}, and \qty{82}{MHz}. The results and the corresponding theoretical predictions are shown in Fig.~\ref{fig:HOM_t}. 
The parameters \(\Gamma_s\) and \(\Gamma_i\) are extracted from the cross-correlation functions measured between Det1 and Det3, and Det1 and Det4. We set \(\kappa_1=0\) and use Eq.~\eqref{d320} together with Eq.~\eqref{d360} to calculate the theoretical prediction. See the Supplementary Material for details. From our results in Fig.~\ref{fig:HOM_t}, one can see that the oscillation frequency of the interference pattern decreases as the frequency detuning between the two independent thermal states diminishes. Finally, it vanishes as detuning approaches zero. Altogether, our results nicely agree with the theoretical predictions. The slightly mismatch between the experimental data and the simulation results mainly lie in experimental imperfection, such as the frequency stability of the pump lasers, bandwidth mismatch, and polarization mismatch.

\begin{figure}[!htbp]
    \centering
    \includegraphics[width=\linewidth]{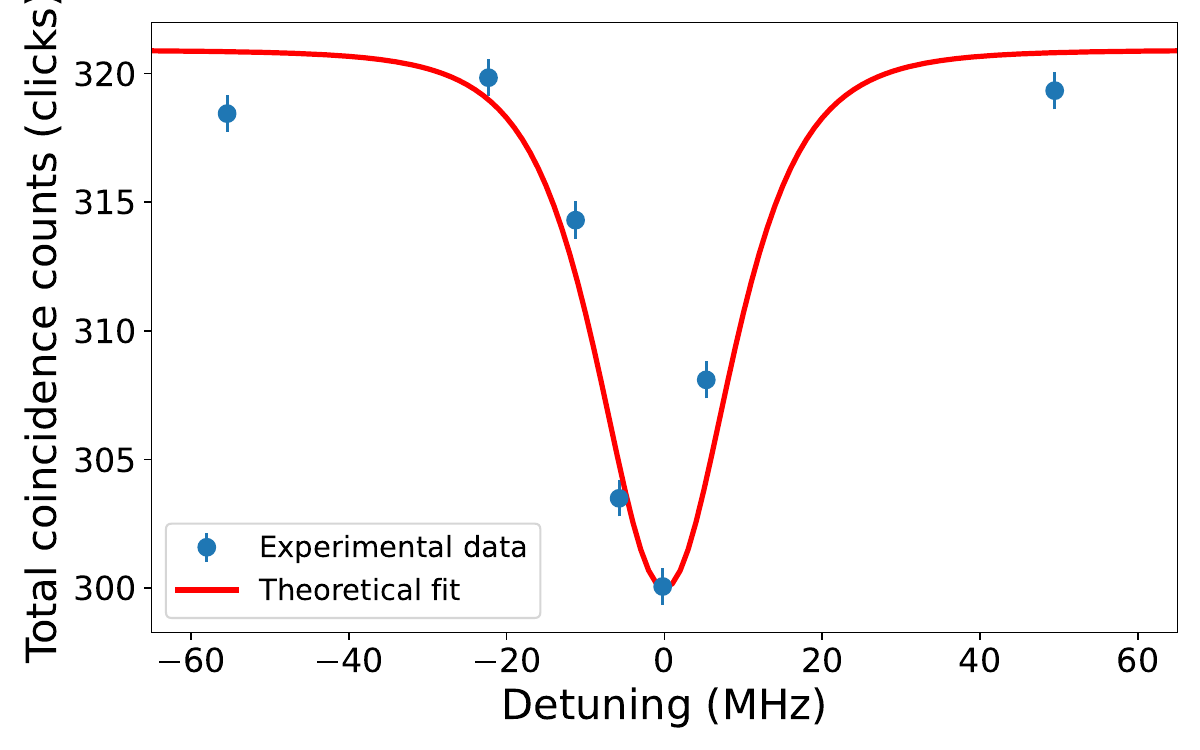}
    \caption{HOM-like interference between two thermal states in frequency domain. The errors are calculated assuming that the main uncertainties come from the photon counting statistics, and all statistics are Poissonian.}
    \label{fig:HOM_f}
\end{figure}

When the frequency detuning is close to zero, the two photons entering from the two input ports tend to exit together from the same output port. As a result, one expects that the total coincidence count between the two output ports will decrease. This behavior can be observed as follows:  for each detuning \(\delta\nu\), the coincidence counts are integrated within the coincidence window \(t\), which is defined as \(\lvert t \rvert \leq 1/(2\uppi \Delta \omega)\), and where \(\Delta \omega\) is the bandwidth of the thermal state. This time interval \(t\) corresponds to the coherence time of the thermal state. The results and the corresponding fit are shown in Fig.~\ref{fig:HOM_f}, from which one can see that the minimum occurs when the frequency difference between the two input states is close to zero.

\begin{figure}[t!]
    \centering
    \includegraphics[width=\linewidth]{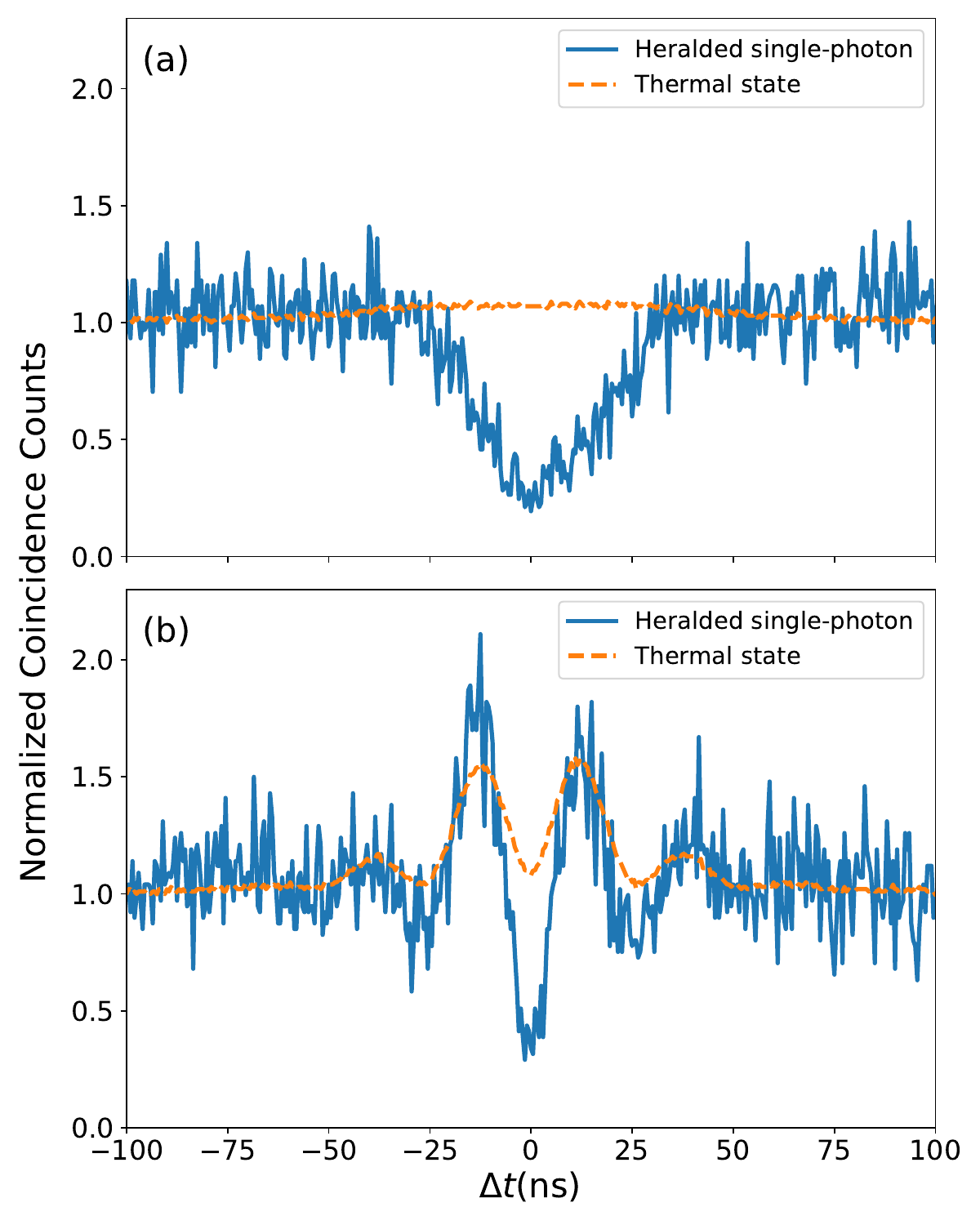}
    \caption{HOM interference between two independent narrowband sources at different frequency detunings as a function of \(\Delta t\). (a) Data measured at zero detuning. (b) Data measured at \qty{37}{\MHz} detuning. For the heralded single-photon results, \(\Delta t\) represents the time difference between the two idlers (measured between Det3 and Det4), while for the thermal state results, it corresponds to $\tau$ defined above. The blue solid lines represent the HOM interference between two heralded single-photon sources, while the orange dashed lines represent the HOM-like interference between two thermal states.}
    \label{fig:HOM_2WGMRs}
\end{figure}

Finally, by measuring 4-fold coincidences we measure the HOM interference between two heralded single-photon sources~\cite{huang2025indistinguishable} under different frequency detunings and compare the results with the HOM-like interference between two thermal states. The results are shown in Fig.~\ref{fig:HOM_2WGMRs}. The HOM interference visibility for two independent heralded single-photon sources is \(76.4\%\pm2.1\%\) at zero detuning and \(63.2\%\pm4.0\%\) at a detuning of \qty{37}{MHz}. 
As seen in the figure, the HOM interference between the two heralded single-photon sources follows the same oscillatory behavior as that between the two thermal states. 

This observation allows us to conclude that it is possible to quickly determine the frequency detuning between two narrowband sources based on thermal state HOM-like interference instead of performing heralded single-photon interference. Measuring a two-fold coincidence typically requires less measurement time to collect sufficient data for statistical purposes, in other words, much more probable, than measuring a four-fold coincidence. For example, in our case, the two-fold coincidence measurement (between the two thermal states) takes only ten minutes, whereas the four-fold coincidence measurement (between the two heralded single-photon sources) takes two hours. Furthermore, we emphasize that this method is particularly useful for the heralded single-photon sources realized via SPDC or four-wave mixing, as the experimental setup for measuring the HOM interference between heralded single-photon sources is identical to that for thermal states. This is due to the fact that photonic source is capable of producing both required states in the same setup. This enables a continuous monitoring of the frequency difference without the need for any realignment of the system.

\section{Conclusion}
In conclusion, we have demonstrated the HOM-like interference between two narrowband thermal states. The experimental results show that a beat signal appears when the two input states are detuned in frequency. The oscillation frequency of the beat signal is directly related to the absolute value of the frequency difference between the two input states and agrees well with the theoretical predictions. The total coincidence count reaches a minimum when the two thermal states are frequency-degenerate, which can be used to determine the zero detuning point. Our results with heralded single photons show in interplay on the boundary between the indistinguishability and distinguishability of non-classical light. In our case, we increase the degree of distinguishability by introducing a frequency detuning, which causes beating in the HOM interference. As the indistinguishability is an important concept behind many quantum optical phenomena, our results show that the spectral degree of freedom has a versatile effect on it. 

This method is useful not only because it provides sufficient frequency precision at the single-photon level, but also because thermal states are usually readily available from existing sources and the measurement setup is identical to a typical HOM interference configuration using heralded single-photons, making it easy to implement. We believe this calibration method will become important for realizing quantum information applications with independent narrowband photonic sources.

\section*{Acknowledgment}
This research was conducted within the scope of the project QuNET, funded by the German Federal Ministry of Science, Technology, and Space (BMFTR) in the context of the federal government’s research framework in IT-security ``Digital. Secure. Sovereign'' (Grant 16KIS1264)

\bibliography{HOM_bib}

\end{document}


\title{Hong--Ou--Mandel-like interference between two independent narrowband thermal states \\ -- Supplementary Material --}

\author{Sheng-Hsuan Huang}
 \email{sheng-hsuan.huang@mpl.mpg.de}
\affiliation{Department of Physics, Friedrich-Alexander-Universität Erlangen-Nürnberg, Staudtstrasse 7/A3, 91058 Erlangen, Germany} 
\affiliation{Max Planck Institute for the Science of Light, Staudtstrasse 2, 91058 Erlangen, Germany}
\author{Thomas Dirmeier}
\affiliation{Department of Physics, Friedrich-Alexander-Universität Erlangen-Nürnberg, Staudtstrasse 7/A3, 91058 Erlangen, Germany} 
\affiliation{Max Planck Institute for the Science of Light, Staudtstrasse 2, 91058 Erlangen, Germany}
\author{Andrea Aiello}
\affiliation{Max Planck Institute for the Science of Light, Staudtstrasse 2, 91058 Erlangen, Germany}
\author{Kaisa Laiho}
\affiliation{German Aerospace Center (DLR e.V.), Institute of Quantum Technologies, Wihelm-Runge-Str. 10, 89081 Ulm, Germany}
\author{\\Dmitry V. Strekalov}
\affiliation{Max Planck Institute for the Science of Light, Staudtstrasse 2, 91058 Erlangen, Germany}
\author{Gerd Leuchs}
\affiliation{Department of Physics, Friedrich-Alexander-Universität Erlangen-Nürnberg, Staudtstrasse 7/A3, 91058 Erlangen, Germany} 
\affiliation{Max Planck Institute for the Science of Light, Staudtstrasse 2, 91058 Erlangen, Germany}
\author{Christoph Marquardt}
\affiliation{Department of Physics, Friedrich-Alexander-Universität Erlangen-Nürnberg, Staudtstrasse 7/A3, 91058 Erlangen, Germany}
\affiliation{Max Planck Institute for the Science of Light, Staudtstrasse 2, 91058 Erlangen, Germany} 
%


\date{\today}

\maketitle

\thispagestyle{empty}


\section{General theory}

We consider a system made of three waves of the electromagnetic field, denoted as the \emph{pump}, the \emph{signal}, and the \emph{idler}. The pump wave is described as a \emph{classical} single-mode field oscillating at frequency $\omega_p$. The signal and the idler waves are represented by multimode \emph{quantum} fields with central frequencies $\omega_s$ and $\omega_i$, respectively. Note that we use the term ``wave'' instead of ``mode'' for the signal and the idler beams, because their amplitudes decay with time, so they are not strictly monochromatic. For this reason they are frequently called ``quasi-modes'' of the electromagnetic field. The WGMR and the fields are sketched in Fig. \ref{fig1} below. 
\begin{figure}[!ht]
  \centering
  \includegraphics[scale=3,clip=false,width=0.5\columnwidth,trim = 0 0 0 0]{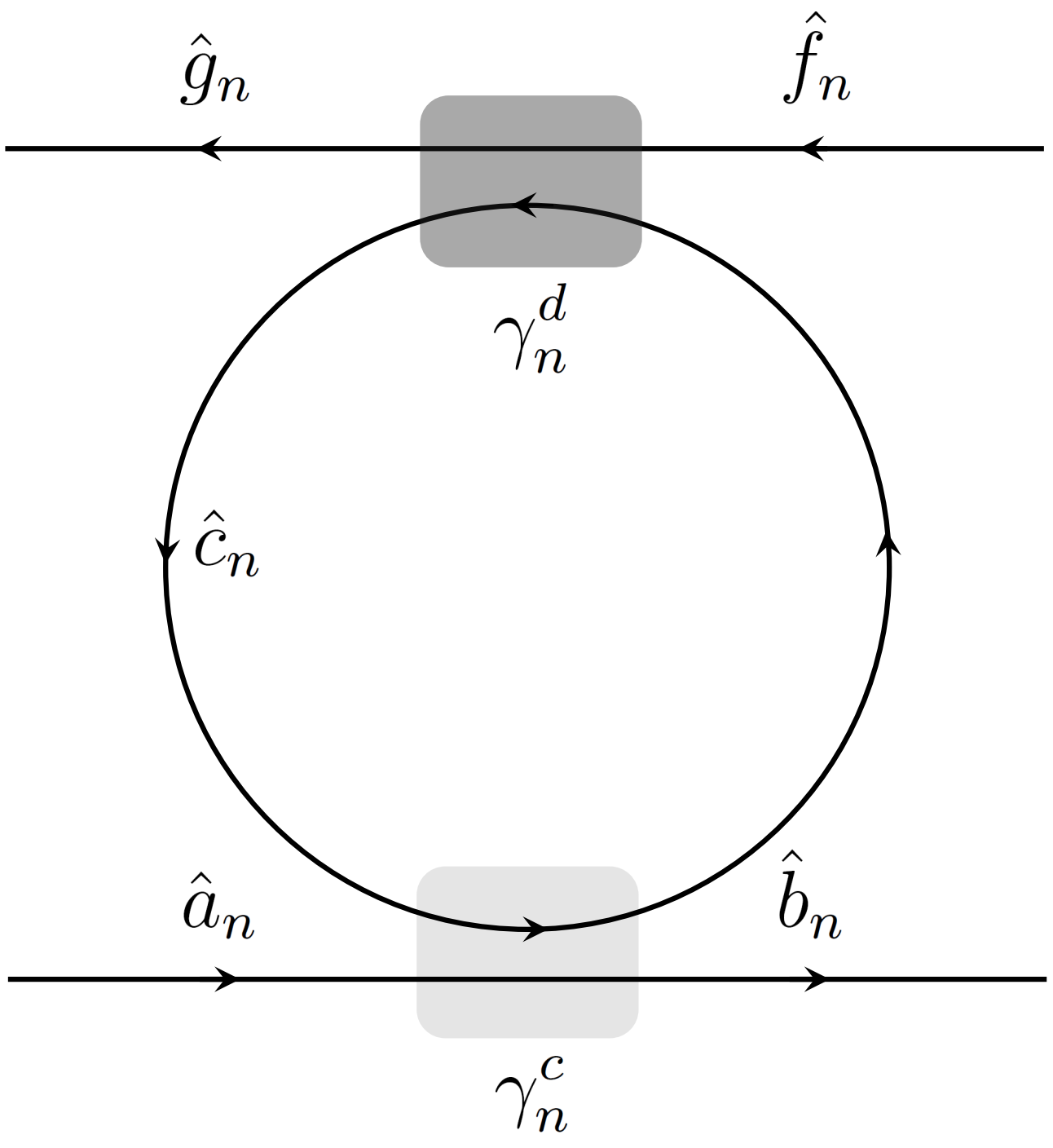}
  \caption{Sketch of the WGMR modeled as a ring cavity. The annihilation operators for the intra-cavity field $\hc_n(t)$, the input and output extra-cavity fields $\ha_n(t), \hb_n(t)$, and the input and output noise fields  $\hf_n(t), \,\hg_n(t)$, are defined in the text, as are the internal losses rate $\gamma_n^d$, and the outcoupling  rate $\gamma_n^c$, where $n=s,i$.}\label{fig1}
\end{figure}
\\
The operators and rates depicted in Fig. \ref{fig1} are defined as follows:
\begin{itemize}
  \item $\hc_n(t)$ is the \emph{intra-cavity} photon-number-amplitude operator, such that the expectation value of $\hc_n^\dagger(t)\hc_n(t)$ gives the average number of photons in the cavity at time $t$. By definition, $\hc_n(t)$ is dimensionless\footnote{Here and hereafter, square brackets denote the dimension of the quantity within them. For example, if $x(t)$ and $v(t)$ denote, respectively, the position and the velocity of a material point,  then $[x(t)] = L$ and $[v(t)] = L T^{-1}$, where $L = L\text{ength}$, and $T = T\text{ime}$.}, that it $\bigl[\hc_n(t)  \bigr] = 1$. By construction, $\hc_n(t)$ and  $\hc_n^\dagger(t)$ do not satisfy  canonical commutation relations \cite{ueda1994anomalous}. 
  \item $\ha_n(t)$ and $\hb_n(t)$ are the \emph{extra-cavity} input and output field operators, respectively, and $\gamma_n^c$ is the coupling rate between internal and external fields, such that 
\begin{align}\label{a10}
\hb_n(t) = - \ha_n(t) +   \sqrt{\gamma_n^c} \, \hc_n(t), \qquad (n=i, s).
\end{align}
By definition, $\ha_n(t)$ and  $\hb_n(t)$ fulfill the canonical commutation relations 
\begin{align}\label{a20}
\bigl[ \ha_n(t), \ha_m^\dagger (t')\bigr] = \delta_{nm}\delta(t-t'), \qquad \text{and} \qquad \bigl[ \hb_n(t), \hb_m^\dagger (t')\bigr] = \delta_{nm} \delta(t-t'), 
\end{align}
where $(n,m = i, s)$. Equations \eqref{a20} imply that  $\ha_n(t)$ and $\hb_n(t)$ have the dimension
$\bigl[\ha_n(t)  \bigr] = T^{-1/2} = \bigl[\hb_n(t)  \bigr] $.
  \item Finally, $\hf_n(t)$ and $\hg_n(t)$ are the extra-cavity \emph{noise}  input and output field operators, respectively, and $\gamma_n^d$ is the coupling rate between the internal field and the noise fields, such that 
\begin{align}\label{a30}
\hg_n(t) = - \hf_n(t) +   \sqrt{\gamma_n^d} \, \hc_n(t), \qquad (n=i, s).
\end{align}
By definition, $\hf_n(t)$ and  $\hg_n(t)$ fulfill the canonical commutation relations 
\begin{align}\label{a40}
\bigl[ \hf_n(t), \hf_m^\dagger (t')\bigr] = \delta_{nm}\delta(t-t'), \qquad \text{and} \qquad \bigl[ \hg_n(t), \hg_m^\dagger (t')\bigr] = \delta_{nm} \delta(t-t'), 
\end{align}
and, consequently, $\bigl[\hf_n(t)  \bigr] = T^{-1/2} = \bigl[\hg_n(t)  \bigr] $.
\end{itemize}
It should be noted that from a mathematical point of view, internal losses and external coupling are handled in exactly the same way in this formalism.

\subsection{Writing the equations of motion}

To find the equations obeyed by the fields defined above, we proceed following chapter  6 of the book by H. Haus \cite{haus2000electromagnetic}, and  the article \cite{shafiee2020nonlinear}. We begin with the ideal case of a closed, lossless cavity with $\gamma^c_n = 0 =\gamma^d_n$, and only later  we introduce positive coupling rates $\gamma^d_n >0 $ and $\gamma^c_n >0$. In a closed, lossless cavity, the number of intracavity photons must be time-independent. Therefore, the multimode field operators $\hc_n(t)$ can be replaced by the single-mode annihilation operators $\hchi_n$, which obeys the canonical commutation relations $\bigl[\hchi_n, \, \hchi_m^\dagger \bigr] = \delta_{nm}$.  The Hamiltonian describing a classical pump wave of frequency $\omega_p$ and complex amplitude $a_p$, interacting with the signal and idler modes of frequencies $\omega_s$ and $\omega_i$, respectively, is given by 
\begin{align}\label{a50}
\hH =  \hbar \omega_s \hchi_s^\dagger \hchi_s + \hbar \omega_i \hchi_i^\dagger \hchi_i 
+ \hbar \kappa \left( a_p \, e^{- i \omega_p t} \hchi_s^\dagger \hchi_i^\dagger + a_p^* \, e^{ i \omega_p t} \hchi_i \hchi_s \right),
\end{align}
where $\kappa$ is the nonlinear coupling rate, with $[ \kappa ] = T^{-1}$. In this equation $\hchi_s, \hchi_i$ and $a_p$ are dimensionless. The Heisenberg equations of motion for the  signal and idler modes are, 
\begin{subequations}\label{a60}
\begin{align}
\frac{d \hchi_s}{d t}  & = \frac{i}{\hbar} \bigl[ \hH, \, \hchi_s \bigr] \nonumber \\[6pt]
 & = - i \omega_s  \hchi_s - i \kappa \, a_p e^{-i \omega_p t}  \hchi_i^\dagger, \label{a60A} \\[6pt]
\frac{d \hchi_i}{d t}  & = \frac{i}{\hbar} \bigl[ \hH, \, \hchi_i \bigr] \nonumber \\[6pt]
 & = - i \omega_i  \hchi_i - i \kappa \, a_p e^{-i \omega_p t}  \hchi_s^\dagger. \label{a60B} 
\end{align}
\end{subequations}
When the losses and the outcoupling are reinstated, we must go back to the multimode field operators $\hc_s(t)$ and $\hc_i(t)$, etc., and make the ansatz,
\begin{subequations}\label{a70}
\begin{align}
\frac{d \hc_s}{d t}  & = - i \omega_s  \hc_s - i \kappa \, a_p e^{-i \omega_p t}  \hc_i^\dagger - \frac{\gamma_s^c + \gamma_s^d}{2}\hc_s + \sqrt{\gamma_s^c} \, \ha_s - \sqrt{\gamma_s^d} \, \hf_s, \label{a70A} \\[6pt]
\frac{d \hc_i}{d t}  & = - i \omega_i  \hc_i - i \kappa \, a_p e^{-i \omega_p t}  \hc_s^\dagger - \frac{\gamma_i^c + \gamma_i^d}{2}\hc_i + \sqrt{\gamma_i^c} \, \ha_i - \sqrt{\gamma_i^d} \, \hf_i . \label{a70B} 
\end{align}
\end{subequations}
These two equations must be supplied with the equations that couple the intra-cavity fields with the extra-cavity ones, as shown by Eqs. \eqref{a10} and \eqref{a30}, that we rewrite here for completeness,
\begin{subequations}\label{a80}
\begin{align}
\hb_n(t) & = - \ha_n(t) +   \sqrt{\gamma_n^c} \, \hc_n(t), \label{a80A} \\[6pt]
\hg_n(t) & = - \hf_n(t) +   \sqrt{\gamma_n^d} \, \hc_n(t), \label{a80B} 
\end{align}
\end{subequations}
where $n=i,s$.
Note that we cannot obtain the lossless part of equations \eqref{a70} by using a Hamiltonian similar to \eqref{a50} with $\hchi_n$ replaced with $\hc_n(t)$, because the intracavity field operators do not obey canonical commutation relations, so we do not know how to calculate the commutator in the Heisenberg equations. This can be seen directly from \eqref{a80}, imposing the validity of the commutation relations \eqref{a20} and \eqref{a40}. A straightforward calculation shows that
\begin{align}\label{a90}
\delta_{nm} \delta (t - t') & = \bigl[ \hb_n(t), \, \hb_m^\dagger(t') \bigr] \nonumber \\[6pt]
 & = \bigl[ - \ha_n(t) +   \sqrt{\gamma_n^c} \, \hc_n(t), \,  - \ha_m^\dagger(t') +   \sqrt{\gamma_m^c} \, \hc_m^\dagger(t') \bigr] \nonumber \\[6pt]
& = \bigl[  \ha_n(t) , \,   \ha_m^\dagger(t')  \bigr] -  \sqrt{\gamma_m^c}  \bigl[ \ha_n(t) , \,  \hc_m^\dagger(t') \bigr] -  \sqrt{\gamma_n^c}  \bigl[ \hc_n(t) , \,  \ha_m^\dagger(t') \bigr]  \nonumber \\[6pt]
& \phantom{=.} + \sqrt{\gamma_n^c \gamma_m^c} \bigl[ \hc_n(t), \, \hc_m^\dagger(t') \bigr] .
\end{align}
This implies
\begin{align}\label{a100}
\bigl[ \hc_n(t), \, \hc_m^\dagger(t') \bigr] =  \frac{1}{\sqrt{\gamma_n^c}}  \bigl[ \ha_n(t) , \,  \hc_m^\dagger(t') \bigr] +   \frac{1}{\sqrt{\gamma_m^c}}  
 \bigl[ \hc_n(t) , \,  \ha_m^\dagger(t') \bigr] .
\end{align}
In exactly the same way we can find,
\begin{align}\label{a110}
\bigl[ \hc_n(t), \, \hc_m^\dagger(t') \bigr] =  \frac{1}{\sqrt{\gamma_n^d}}  \bigl[ \hf_n(t) , \,  \hc_m^\dagger(t') \bigr] +   \frac{1}{\sqrt{\gamma_m^d}}  
 \bigl[ \hc_n(t) , \,  \hf_m^\dagger(t') \bigr] .
\end{align}
In the last two equations it should be noticed the complete analogy between the source terms $\ha_n(t)/\sqrt{\gamma_n^c}$ and the noise terms $\hf_n(t)/\sqrt{\gamma_n^d}$. To calculate the commutators above, we must first solve the equations \eqref{a70}.

\subsection{Solving the equations of motion}

To solve the set of first-order differential equations \eqref{a70}, it is convenient to use the Fourier transform formalism \cite{OsgoodBook}. We adopt the following convention (here illustrated for the generic Fourier pair of operators $\left\{ \ha(t), \, \ha(\omega)\right\}$):
\begin{subequations}\label{a120}
\begin{align}
\ha(t) & = \frac{1}{\sqrt{2 \pi}} \int_\mathbb{R} \ha(\omega) e^{- i \omega t} \di \omega, \label{a120A} \\[6pt]
\ha(\omega) & = \frac{1}{\sqrt{2 \pi}} \int_\mathbb{R} \ha(t) e^{ i \omega t} \di t. \label{a120B} 
\end{align}
\end{subequations}
This convention implies that
\begin{align}\label{a130}
e^{-i \omega_p t}  \ha^\dagger(t) & = \frac{1}{\sqrt{2 \pi}} \int_\mathbb{R} \ha^\dagger (\omega) e^{ i (\omega - \omega_p) t} \di \omega \nonumber \\[6pt]
& = \frac{1}{\sqrt{2 \pi}} \int_\mathbb{R} \ha^\dagger (\omega_p - \omega') e^{- i \omega'  t} \di \omega', \nonumber \\[6pt]
& = \frac{1}{\sqrt{2 \pi}} \int_\mathbb{R} \ha^\dagger (\omega_p - \omega) e^{- i \omega  t} \di \omega,
\end{align}
where to get the second line we made the change of variable $\omega - \omega_p \to -\omega'$.

A straightforward calculation shows that applying Fourier transform to \eqref{a70}, we get the following algebraic system in four dimensions,
\begin{multline}\label{a140}
\begin{bmatrix}
  v_s(\omega) & 0 & 0 & i \kappa a_p \\[4pt]
  0 & v_s^*(\omega_p - \omega) & -i \kappa a_p^* & 0 \\[4pt]
  0 & i \kappa a_p & v_i(\omega) & 0 \\[4pt]
  -i \kappa a_p^* & 0 & 0 & v_i^*(\omega_p - \omega) \\[4pt]
\end{bmatrix}
\begin{bmatrix}
  \hc_s(\omega) \\[4pt]
  \hc_s^\dagger(\omega_p - \omega) \\[4pt]
  \hc_i(\omega) \\[4pt]
  \hc_i^\dagger(\omega_p - \omega) \\[4pt]
\end{bmatrix}
\\[6pt]
= %
\begin{bmatrix}
  \sqrt{\gamma_s^c} & 0 & 0 & 0 \\[4pt]
  0 & \sqrt{\gamma_s^c} & 0 & 0 \\[4pt]
  0 & 0 & \sqrt{\gamma_i^c} & 0 \\[4pt]
 0 & 0 & 0 & \sqrt{\gamma_i^c} \\[4pt]
\end{bmatrix}
\begin{bmatrix}
  \ha_s(\omega) \\[4pt]
  \ha_s^\dagger(\omega_p - \omega) \\[4pt]
  \ha_i(\omega) \\[4pt]
  \ha_i^\dagger(\omega_p - \omega) \\[4pt]
\end{bmatrix}
-
\begin{bmatrix}
  \sqrt{\gamma_s^d} & 0 & 0 & 0 \\[4pt]
  0 & \sqrt{\gamma_s^d} & 0 & 0 \\[4pt]
  0 & 0 & \sqrt{\gamma_i^d} & 0 \\[4pt]
 0 & 0 & 0 & \sqrt{\gamma_i^d} \\[4pt]
\end{bmatrix}
\begin{bmatrix}
  \hf_s(\omega) \\[4pt]
  \hf_s^\dagger(\omega_p - \omega) \\[4pt]
  \hf_i(\omega) \\[4pt]
  \hf_i^\dagger(\omega_p - \omega) \\[4pt]
\end{bmatrix},
\end{multline}
where we have defined the function
\begin{align}\label{a150}
v_n(\omega) \deff \frac{\Gamma_n}{2} - i \left( \omega- \omega_n \right), \qquad (n=i, s),
\end{align}
with
\begin{align}\label{a210}
\Gamma_n \deff \gamma_n^c + \gamma_n^d , \qquad (n=i, s).
\end{align}
A simple matrix inversion permits us to solve \eqref{a140}, to obtain
\begin{subequations}\label{a160}
\begin{align}
\hc_s(\omega) & = \frac{v_i^*(\omega_p-\omega)\sqrt{\gamma_s^c} }{D_s(\omega)} \, \ha_s(\omega)   - i \frac{\kappa \, a_p \, \sqrt{\gamma_i^c}}{D_s(\omega)} \, \ha^\dagger_i(\omega_p - \omega) \nonumber \\[8pt]
& \phantom{=.} - \frac{v_i^*(\omega_p-\omega)\sqrt{\gamma_s^d} }{D_s(\omega)}  \hf_s(\omega) + i \frac{\kappa \, a_p \, \sqrt{\gamma_i^d}}{D_s(\omega)} \, \hf^\dagger_i(\omega_p - \omega)  , \label{a160A} \\[10pt]
\hc_i(\omega) & =  \frac{v_s^*(\omega_p-\omega)\sqrt{\gamma_i^c} }{D_i(\omega)} \, \ha_i(\omega)   - i \frac{\kappa \, a_p \, \sqrt{\gamma_s^c}}{D_i(\omega)} \, \ha^\dagger_s(\omega_p - \omega) \nonumber \\[8pt]
& \phantom{=.} - \frac{v_s^*(\omega_p-\omega)\sqrt{\gamma_i^d} }{D_i(\omega)}  \hf_i(\omega) + i \frac{\kappa \, a_p \, \sqrt{\gamma_s^d}}{D_i(\omega)} \, \hf^\dagger_s(\omega_p - \omega)  ,  \label{a160B} 
\end{align}
\end{subequations}
where we have defined the denominators $D_s(\omega)$ and $D_i(\omega)$, as
\begin{subequations}\label{a170}
\begin{align}
D_s(\omega) & \deff v_s(\omega) v^*_i(\omega_p-\omega) - \kappa^2 \abs{a_p}^2,\label{a170A} \\[6pt]
D_i(\omega) & \deff v^*_s(\omega_p-\omega) v_i(\omega) - \kappa^2 \abs{a_p}^2.  \label{a170B} 
\end{align}
\end{subequations}

Now we are ready to verify Eqs. \eqref{a100} and \eqref{a110}, which read, in the Fourier space, 
\begin{align}\label{a180}
\bigl[ \hc_n(\omega), \, \hc_m^\dagger(\omega') \bigr] =  \frac{1}{\sqrt{\gamma_n^c}}  \bigl[ \ha_n(\omega) , \,  \hc_m^\dagger(\omega') \bigr] +   \frac{1}{\sqrt{\gamma_m^c}}  
 \bigl[ \hc_n(\omega) , \,  \ha_m^\dagger(\omega') \bigr] ,
\end{align}
and 
\begin{align}\label{a190}
\bigl[ \hc_n(\omega), \, \hc_m^\dagger(\omega') \bigr] =  \frac{1}{\sqrt{\gamma_n^d}}  \bigl[ \hf_n(\omega) , \,  \hc_m^\dagger(\omega') \bigr] +   \frac{1}{\sqrt{\gamma_m^d}}  
 \bigl[ \hc_n(\omega) , \,  \hf_m^\dagger(\omega') \bigr] ,
\end{align}
with $(n,m=i, s)$. For example, a direct calculation gives
\begin{subequations}\label{a200}
\begin{align}
\bigl[ \hc_s(\omega), \, \hc_s^\dagger(\omega') \bigr] & = \delta(\omega-\omega') \, \frac{\abs{v_i(\omega_p - \omega)}^2 \Gamma_s - \kappa^2 \abs{a_p}^2 \Gamma_i}{\abs{D_s(\omega)}^2} , \label{a200A} \\[8pt]
\bigl[ \hc_i(\omega), \, \hc_i^\dagger(\omega') \bigr] & = \delta(\omega-\omega') \, \frac{\abs{v_s(\omega_p - \omega)}^2 \Gamma_i - \kappa^2 \abs{a_p}^2 \Gamma_s}{\abs{D_i(\omega)}^2} .  \label{a200B} 
\end{align}
\end{subequations}
Similarly, we get 
\begin{subequations}\label{a220}
\begin{align}
\frac{1}{\sqrt{\gamma_s^c}}\bigl[ \ha_s(\omega), \, \hc_s^\dagger(\omega') \bigr]  & = \delta(\omega-\omega')\frac{v_i(\omega_p - \omega)}{D_s^*(\omega)} , \label{a220A} \\[8pt]
\frac{1}{\sqrt{\gamma_i^c}}\bigl[ \ha_i(\omega), \, \hc_i^\dagger(\omega') \bigr] & = \delta(\omega-\omega')\frac{v_s(\omega_p - \omega)}{D_i^*(\omega)}. \label{a220B} 
\end{align}
\end{subequations}
Using \eqref{a200} and \eqref{a220} it is not difficult to see that \eqref{a180} with $n=s=m$, is satisfied. In the same way, we can verify the other equalities.

Using \eqref{a80} we can write the output fields as,
\begin{subequations}\label{a230}
\begin{align}
\hb_n(\omega) & = - \ha_n(\omega) +   \sqrt{\gamma_n^c} \, \hc_n(\omega), \label{a230A} \\[6pt]
\hg_n(\omega) & = - \hf_n(\omega) +   \sqrt{\gamma_n^d} \, \hc_n(\omega), \label{a230B} 
\end{align}
\end{subequations}
where $n=i,s$. Finally, substituting  \eqref{a160} into \eqref{a230A}, we obtain
\begin{subequations}\label{z160}
\begin{align}
\hb_s(\omega) & = A_{ss}(\omega)  \ha_s(\omega)   +  A_{si}(\omega)  \ha^\dagger_i(\omega_p - \omega) \nonumber \\[8pt]
& \phantom{=.} + F_{ss}(\omega)  \hf_s(\omega)  + F_{si}(\omega)   \hf^\dagger_i(\omega_p - \omega)  , \label{z160A} \\[10pt]
\hb_i(\omega) & = A_{ii}(\omega) \ha_i(\omega) + A_{is}(\omega) \ha^\dagger_s(\omega_p - \omega) \nonumber \\[8pt]
& \phantom{=.} +F_{ii}(\omega)  \hf_i(\omega) + F_{is}(\omega) \hf^\dagger_s(\omega_p - \omega)  ,  \label{z160B} 
\end{align}
\end{subequations}
where we have defined
\begin{subequations}\label{z150}
\begin{align}
A_{ss}(\omega) & = \frac{\bigl[\Gamma_i/2  - i \left(\omega - \omega_s - \Delta \right)\bigr] \bigl[\gamma^c_s - \Gamma_s/2  + i \left(\omega - \omega_s \right)\bigr] + \abs{\kappa_1}^2}{\bigl[\Gamma_i/2  - i \left(\omega - \omega_s - \Delta \right)\bigr] \bigl[\Gamma_s/2  - i \left(\omega - \omega_s \right)\bigr] - \abs{\kappa_1}^2}  , \label{z150A} \\[8pt]
A_{si}(\omega)& = \frac{- i \, \kappa_1  \sqrt{\gamma^c_s \gamma^c_i}}{\bigl[\Gamma_i/2  - i \left(\omega - \omega_s - \Delta \right)\bigr] \bigl[\Gamma_s/2  - i \left(\omega - \omega_s \right)\bigr] - \abs{\kappa_1}^2} , \label{z150B} \\[10pt]
F_{ss}(\omega) & = \frac{-\bigl[\Gamma_i/2  - i \left(\omega - \omega_s - \Delta \right)\bigr] \sqrt{\gamma^c_s \gamma^d_s} }{\bigl[\Gamma_i/2  - i \left(\omega - \omega_s - \Delta \right)\bigr] \bigl[\Gamma_s/2  - i \left(\omega - \omega_s \right)\bigr] - \abs{\kappa_1}^2} , \label{z150C} \\[8pt]
F_{si}(\omega)&  = \frac{- i \, \kappa_1  \sqrt{\gamma^c_s \gamma^d_i}}{\bigl[\Gamma_i/2  - i \left(\omega - \omega_s - \Delta \right)\bigr] \bigl[\Gamma_s/2  - i \left(\omega - \omega_s \right)\bigr] - \abs{\kappa_1}^2} ,  \label{z150D} 
\end{align}
\end{subequations}    
with 
\begin{align}\label{kappa1}
\kappa_1 \deff \kappa \,a_p,
\end{align}
and $\{ A_{ii}, A_{is}, F_{ii}, F_{is} \}$ are obtained by exchanging everywhere the indices $s$ and $i$ in the formulas \eqref{a150} for  $\{ A_{ss}, A_{si}, F_{ss}, F_{si} \}$, respectively.  

\section{Second-order correlation functions}

\begin{figure}[!ht]
  \centering
  \includegraphics[scale=3,clip=false,width=0.5\columnwidth,trim = 0 0 0 0]{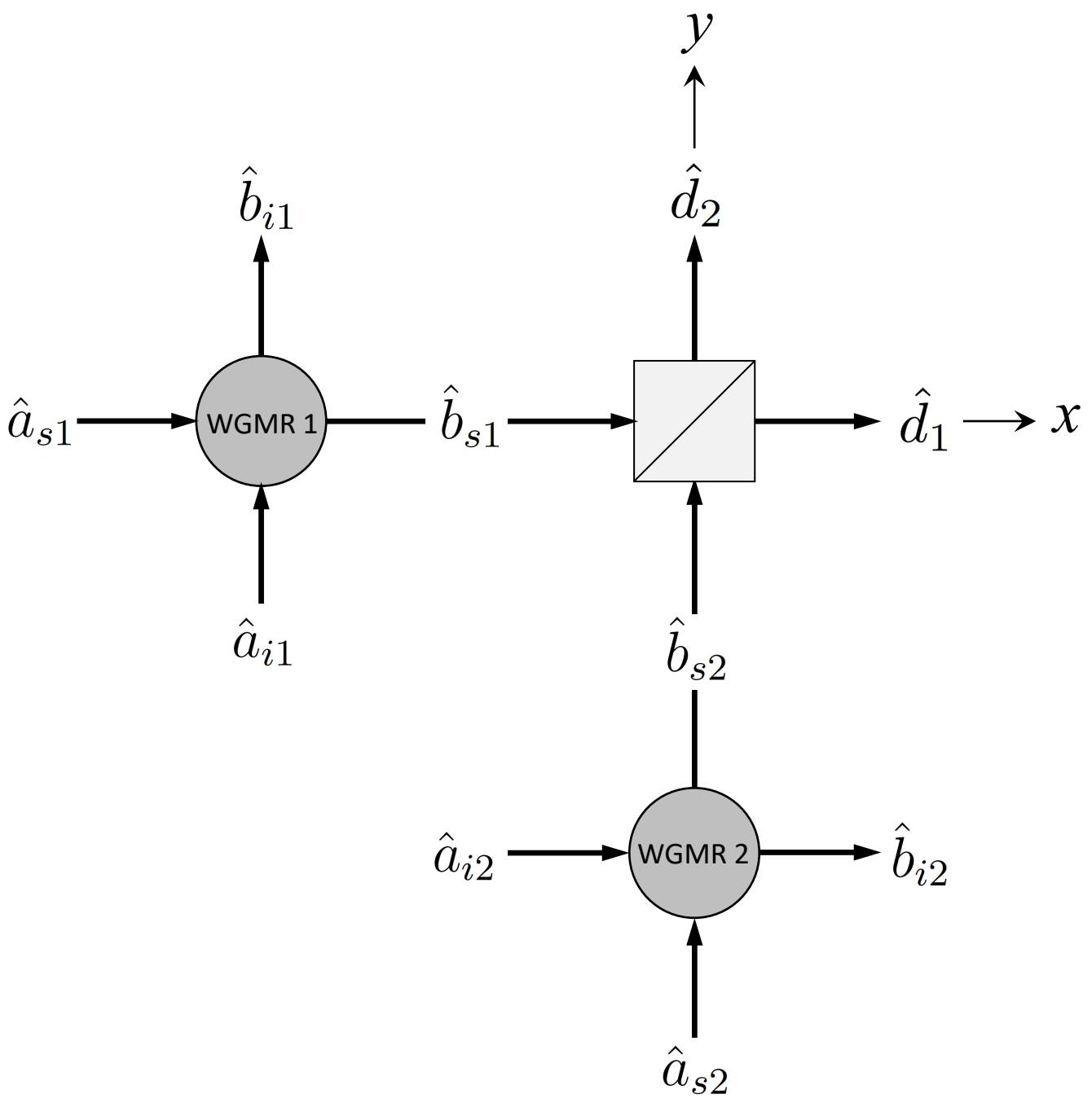}
  \caption{Sketch of the theoretical model of the experimental setup. The ``signal'' beams from two independent WGMRs enters the  ports ``$1$'' and ``$2$'' of a beam splitter (BS), which are  aligned, respectively, to the axes $x$ and $y$ of a Cartesian reference frame. The ``idler'' photons are not detected.}\label{fig2}
\end{figure}

Let us consider now the experimental scenario schematically depicted in Fig. \ref{fig2}.
Two independent  thermal states carried by the signal beams   emitted at  frequencies $\omega_{s1}$ and $\omega_{s2}$ by the two WGMRs,  enter the  input ports of a beam splitter (BS). The latter is characterized by the transmission and reflection coefficients $\mT$ and $\maR$, respectively, with the standard properties:
\begin{align}\label{d10}
\abs{\mT}^2 + \abs{\maR}^2 = 1, \qquad \text{and} \qquad \mT \maR^* +  \mT^* \maR = 0.
\end{align}
The BS transforms the annihilation operators $\hb_{s1}(t)$ and $\hb_{s2}(t)$ of the input field into the annihilation operators $\hd_1(t)$ and $\hd_2(t)$ of the output field, according to
\begin{subequations}\label{d20}
\begin{align}
\hd_1(t) & = \mT \, \hb_{s1}(t) + \maR \hb_{s2}(t), \label{d20A} \\[6pt]
\hd_2(t) & = \maR \hb_{s1}(t) + \mT \,  \hb_{s2}(t), \label{d20B}
\end{align}
\end{subequations}
where
\begin{align}\label{d22}
\hb_{s \alpha}(t)  = \frac{1}{\sqrt{2 \pi}} \int_\mathbb{R} \hb_{s \alpha}(\omega) e^{- i \omega t} \di \omega, \quad (\alpha =1,2),
\end{align}
with $\hb_{s \alpha}(\omega)$ being given by \eqref{a160} and \eqref{a150}, where all indexes ``$p$'', ``$i$'' and  ``$s$'' are replaced by ``$p \alpha$'', ``$i \alpha$'' and  ``$s \alpha$'', respectively, and $\Delta \to \Delta_\alpha = \omega_{p\alpha} - \omega_{s\alpha} - \omega_{i \alpha}$, where $\alpha=1,2$ is the beam index.
In our experiment we have zero detuning in both WGMRs, so that $\Delta_1= 0 = \Delta_2$. Moreover, because of the small frequency detuning 
\begin{align}\label{d24}
\delta \omega \deff \omega_{s2} - \omega_{s1},
\end{align}
between the two thermal sources,  the loss and outcoupling rates are  basically frequency-independent, so that $\{\gamma^c_{s \alpha}, \gamma^c_{i\alpha}, \gamma^d_{s\alpha}, \gamma^d_{i\alpha} \}=\{\gamma^c_s, \gamma^c_i, \gamma^d_s, \gamma^d_i \}$, and $\kappa_{1\alpha} = \kappa_1$.

Two detectors, labelled $D_1$ and $D_2$, are placed after the BS on the axes $x$ and $y$, respectively, at  $x = x_1>0 $ and $y= y_2>0$, where the origin of the coordinates $x=0=y$ is taken at the center of the BS. We  calculate the \emph{interbeam} degree of second-order coherence measured by the detectors $D_1$ and $D_2$, as
\begin{align}\label{d30}
g^{(2)} \left( \tau \right) = \frac{\langle \hd_{1}^\dagger(\tau_1) \hd_{2}^\dagger(\tau_2) \hd_2(\tau_2)\hd_1(\tau_1) \rangle}{\langle \hd_{1}^\dagger(\tau_1) \hd_1(\tau_1) \rangle \langle \hd_{2}^\dagger(\tau_2) \hd_2(\tau_2) \rangle} ,
\end{align}
where  the field operators in the equation above are evaluated  at different times $t_1$ and $t_2$,  and we have defined
\begin{align}\label{d32}
\tau_1 & = t_1 - \frac{x_1}{c}, \nonumber \\[6pt]
\tau_2 & = t_2 - \frac{y_2}{c},
\end{align}
and
\begin{align}\label{d34}
\tau & = \tau_2 - \tau_1, \nonumber \\[6pt]
 & = t_2 - t_1 - \frac{y_2 - x_1}{c},
\end{align}
Now we will calculate first the two terms in the denominator, and then the big term in the numerator.

\subsubsection{Calculation of $\langle  \hd^\dagger_{\mu}(\tau_1)  \hd_{\nu}(\tau_2)  \rangle$.}

To perform this calculation it is convenient to rewrite \eqref{d20} as 
\begin{align}\label{d40}
\hd_\mu(t) = \sum_{\alpha=1}^2 S_{\mu \alpha} \hb_{s \alpha}(t), \qquad (\mu=1,2),
\end{align}
where we have defined the $2 \times 2$ matrix $S$, as
\begin{align}\label{d50}
S \deff \begin{bmatrix}
          S_{11} & S_{12} \\
          S_{21} & S_{22} \\
        \end{bmatrix}
= \begin{bmatrix}
          \mT & \maR \\
          \maR & \mT \\
        \end{bmatrix}.
\end{align}
Then, we can write
\begin{align}\label{d60}
\langle  \hdd_{\mu}(\tau_1)  \hd_{\nu}(\tau_2)  \rangle = \sum_{\alpha, \beta=1}^2 S_{\mu \alpha}^* S_{\nu \beta} \langle  \hbd_{s \alpha}(\tau_1)  \hb_{s \beta}(\tau_2)  \rangle, 
\end{align}
where
\begin{align}\label{d70}
 \langle  \hbd_{s\alpha}(\tau_1)  \hb_{s\beta}(\tau_2)  \rangle  = \frac{1}{2 \pi} \int_{\mathbb{R}^2}  e^{i \omega_1 \tau_1 - i \omega_2 \tau_2}  \langle  \hbd_{s\alpha}(\omega_1)  \hb_{s\beta}(\omega_2)  \rangle  \di \omega_1 \di \omega_2 .
\end{align}
From \eqref{z160A} we have that
\begin{align}\label{d80}
\langle  \hbd_{s \alpha}(\omega_1)  \hb_{s \beta}(\omega_2)  \rangle & = \bra{0} \left[ \cancel{A_{s s \alpha}^*(\omega_1)   \ha_{s \alpha}^\dagger(\omega_1)} + \textcolor{blue}{A_{s i \alpha}^*(\omega_1)  \ha_{i \alpha}(\omega_{p \alpha} - \omega_1)} \right. \nonumber \\[6pt]
& \phantom{= \bra{0} [].} \left. +\cancel{ F_{s s \alpha}^*(\omega_1)   \hf_{s \alpha}^\dagger(\omega_1)} + F_{s i \alpha}^*(\omega_1)  \hf_{i \alpha}(\omega_{p \alpha} - \omega_1)\right] \nonumber \\[6pt]
& \phantom{=..} \times \left[ \cancel{A_{s s \beta}(\omega_2)   \ha_{s \beta}(\omega_2)} + \textcolor{blue}{A_{s i \beta}(\omega_2)  \ha^\dagger_{i \beta}(\omega_{p \beta} - \omega_2)} \right. \nonumber \\[6pt]
& \phantom{= \bra{0} [].} \left. + \cancel{ F_{s s \beta}(\omega_2)   \hf_{s \beta}(\omega_2) } + F_{s i \beta}(\omega_2)  \hf^\dagger_{i \beta}(\omega_{p \beta} - \omega_2)\right] \ket{0} \nonumber \\[6pt]
& = A_{s i \alpha}^*(\omega_1)A_{s i \beta}(\omega_2) \mean{0}{\ha_{i \alpha}(\omega_{p \alpha} - \omega_1)\ha^\dagger_{i \beta}(\omega_{p \beta} - \omega_2)}{0} \nonumber \\[6pt]
& \phantom{=.} + F_{s i \alpha}^*(\omega_1)F_{s i \beta}(\omega_2) \mean{0}{\hf_{i \alpha}(\omega_{p \alpha} - \omega_1)\hf^\dagger_{i \beta}(\omega_{p \beta} - \omega_2)}{0}\nonumber \\[6pt]
& = \delta_{\alpha \beta} \delta \left( \omega_1 - \omega_2 \right) \left[\abs{A_{s i \alpha}(\omega_1)}^2  + \abs{F_{s i \alpha}(\omega_1)}^2 \right].
\end{align}
Using this result in \eqref{d70}, we readily obtain
\begin{align}\label{d90}
 \langle  \hbd_{s \alpha}(\tau_1)  \hb_{s \beta}(\tau_2)  \rangle & = \delta_{\alpha \beta} \left\{\frac{1}{2 \pi} \int_{\mathbb{R}} e^{-i \omega \tau} \abs{A_{s i \alpha}(\omega)}^2 \di \omega + \frac{1}{2 \pi} \int_{\mathbb{R}} e^{-i \omega \tau} \abs{F_{s i \alpha}(\omega)}^2  \di \omega \right\} \nonumber \\[6pt]
& \deff  \delta_{\alpha \beta} \bigl\{ \varphi_{A_\alpha}(\tau) + \varphi_{F_\alpha}(\tau) \bigr\},
\end{align}
where we have defined 
\begin{equation}\label{d100}
\begin{split}
 \varphi_{A_\alpha}(\tau) & \deff  \frac{1}{2 \pi}\int_{\mathbb{R}} e^{-i \omega \tau} \abs{A_{s i \alpha}(\omega)}^2 \di \omega ,  \\[6pt]
\varphi_{F_\alpha}(\tau) &  \deff \frac{1}{2 \pi} \int_{\mathbb{R}} e^{-i \omega \tau} \abs{F_{s i \alpha}(\omega)}^2  \di \omega.
\end{split}
\end{equation}
Since the sum $\varphi_{A_\alpha}(\tau) + \varphi_{F_\alpha}(\tau) $ will occur many times, we give a special name to it, that is 
\begin{align}\label{d110}
\phi_\alpha(\tau) \deff  \varphi_{A_\alpha}(\tau) + \varphi_{F_\alpha}(\tau) , \qquad (\alpha = 1,2).
\end{align}
From \eqref{z150}, it follows that
\begin{align}\label{d112}
\abs{A_{s i \alpha}(\omega)}^2 + \abs{F_{s i \alpha}(\omega)}^2 & =  \abs{A_{s i \alpha}(\omega)}^2 \frac{\Gamma_{i}}{\gamma^c_{i }} \nonumber \\[6pt]
& = \frac{\abs{\kappa_{1\alpha}}^2 \gamma^c_{s } \, \Gamma_{i}}{\abs{\bigl[\Gamma_{i}/2  - i \left(\omega - \omega_{s\alpha} \right)\bigr] \bigl[\Gamma_{s}/2  - i \left(\omega - \omega_{s \alpha} \right)\bigr] - \abs{\kappa_{1}}^2}^2}.
\end{align}
Using these results we can rewrite
\begin{align}\label{d115}
\phi_\alpha(\tau) =  \bigl( \abs{\kappa_{1}}^2 \, \gamma^c_{s} \, \Gamma_{i} \bigr) e^{-i \omega_{s \alpha} \tau}I_\alpha(\tau),
\end{align}
where 
\begin{align}\label{d330b}
I_\alpha( \tau )  = \frac{1}{2 \pi}\int_\mathbb{R} \frac{e^{- i  x \tau}}{\abs{\left(\Gamma_{i} /2 - i  x \right) \left(\Gamma_{s} /2 - i x \right) - \abs{\kappa_{1}}^2}^2} \, \di x \,.
\end{align}
A direct calculation with Mathematica  \cite{Mathematica}, gives
\begin{align}\label{d340b}
\frac{I(\tau)}{I(0)} = \frac{a_+ \exp\left(-\dfrac{a_-}{2} \abs{\tau}\right)  - a_- \exp\left(-\dfrac{a_+}{2} \abs{\tau}\right)}{a_+ - a_-},
\end{align}
where
\begin{align}\label{d350b}
a_\pm = \sqrt{\frac{\Gamma_{i}^2 + \Gamma_{s }^2 + 8  \abs{\kappa_1}^2}{2}  \pm \frac{\left( \Gamma_{i} + \Gamma_{s } \right)  
\sqrt{\left( \Gamma_{i} - \Gamma_{s } \right)^2 + 16 \abs{\kappa_1}^2}}{2} }.
\end{align}
When $\Gamma_i \Gamma_s \gg 4 \abs{\kappa_1}^2 $, we can set $\abs{\kappa_1} =0$ and \eqref{d340b} reduces to
\begin{align}\label{d360b}
\frac{I(\tau)}{I(0)} = \frac{\Gamma_i \, e^{ - \Gamma_s \abs{\tau}/2 } -  \Gamma_s \, e^{ - \Gamma_i \abs{\tau}/2 } }{\Gamma_i - \Gamma_s}.
\end{align}

Now, substituting \eqref{d90} into \eqref{d60}, we obtain
\begin{align}\label{d120}
\langle  \hdd_{\mu}(\tau_1)  \hd_{\nu}(\tau_2)  \rangle & = \sum_{\alpha, \beta=1}^2 S_{\mu \alpha}^* S_{\mu \beta}  \delta_{\alpha \beta} \phi_\alpha(\tau) \nonumber \\[6pt]
 & = \sum_{\alpha=1}^2 S_{\mu \alpha}^* S_{\nu \alpha}   \phi_\alpha(\tau).
\end{align}
Using this result, we can rewrite the denominator in \eqref{d30}, as
\begin{align}\label{d130}
\langle \hdd_{1}(\tau_1) \hd_1(\tau_1) \rangle \langle \hdd_{2}(\tau_2) \hd_2(\tau_2) \rangle & = \left[ \sum_{\alpha=1}^2 \abs{S_{1 \alpha}}^2 \phi_\alpha(0) \right] \left[ \sum_{\beta=1}^2 \abs{S_{2 \beta}}^2 \phi_\beta(0) \right] \nonumber \\[6pt]
& = \Bigl[ \abs{\mT}^2 \phi_1(0) + \abs{\maR}^2 \phi_2(0) \Bigr]  \Bigl[ \abs{\maR}^2 \phi_1(0) + \abs{\mT}^2 \phi_2(0) \Bigr].
\end{align}

\subsubsection{Calculation of $\langle \hdd_{s1}(\tau_1) \hdd_{s1}(\tau_2) \hd_{s1}(\tau_2) \hd_{s1}(\tau_1) \rangle$.}

Now, we calculate
\begin{multline}\label{d140}
\langle \hdd_{s1}(\tau_1) \hdd_{s1}(\tau_2) \hd_{s1}(\tau_2) \hd_{s1}(\tau_1) \rangle \\[6pt]
= \frac{1}{(2 \pi)^2}\int_{\mathbb{R}^4} e^{i \tau_1(\omega_1 - \omega_1')  + i\tau_2 (\omega_2 - \omega_2'  )} \langle  \hdd_{s1}(\omega_1')  \hdd_{s1}(\omega_2') \hd_{s1}(\omega_2)  \hd_{s1}(\omega_1) \rangle \di \omega_1' \di \omega_2' \di \omega_1 \di \omega_2.
\end{multline}
To calculate this quantity, it is convenient to write
\begin{align}\label{d150}
\hd_{s1}(\omega_2)  \hd_{s1}(\omega_1)  & = \left[\maR \hb_{s1}(\omega_2) + \mT \hb_{s2}(\omega_2) \right] \left[\mT \hb_{s1}(\omega_1) + \maR \hb_{s2}(\omega_1) \right] \nonumber \\[6pt]
 & = \maR \mT \hb_{s1}(\omega_2) \hb_{s1}(\omega_1)\ket{0} + \mathcal{R}^2  \hb_{s1}(\omega_2) \hb_{s2}(\omega_1) \nonumber \\[6pt]
 & \phantom{=.} \mT^2 \hb_{s2}(\omega_2) \hb_{s1}(\omega_1) + \mT \maR \hb_{s2}(\omega_2) \hb_{s2}(\omega_1) \nonumber \\
 & \deff \sum_{\alpha,\beta=1}^2 G_{\alpha \beta}  \hb_{s\alpha}(\omega_2) \hb_{s\beta}(\omega_1),
\end{align}
where we have defined 
\begin{align}\label{d160}
G \deff \begin{bmatrix}
          G_{11} & G_{12} \\
          G_{21} & G_{22} \\
        \end{bmatrix}
= \begin{bmatrix}
          \maR \mT & \mathcal{R}^2 \\
          \mT^2 & \maR \mT \\
        \end{bmatrix}.
\end{align}
Using \eqref{d150} we can rewrite $\langle  \hdd_{s1}(\omega_1')  \hdd_{s1}(\omega_2') \hd_{s1}(\omega_2)  \hd_{s1}(\omega_1) \rangle$  as,
\begin{align}\label{d170}
\langle  \hdd_{s1}(\omega_1')  \hdd_{s1}(\omega_2') \hd_{s1}(\omega_2)  \hd_{s1}(\omega_1) \rangle =
 \sum_{\alpha',\beta'=1}^2  \sum_{\alpha,\beta=1}^2 G_{\alpha' \beta'}^* G_{\alpha \beta} \langle \hbd_{s\beta'}(\omega_1') \hbd_{s\alpha'}(\omega_2') \hb_{s\alpha}(\omega_2) \hb_{s\beta}(\omega_1) \rangle,
\end{align}
where 
\begin{align}\label{d180}
\langle \hbd_{s\beta'}(\omega_1') \hbd_{s\alpha'}(\omega_2') \hb_{s\alpha}(\omega_2) \hb_{s\beta}(\omega_1) \rangle & = \bra{0} \left[ A_{si   \beta'}^*(\omega_1')  \ha_{i 1}(\omega_{p \beta'} - \omega_1') + F_{s i \beta'}^*(\omega_1')  \hf_{i \beta'}(\omega_{p \beta'} - \omega_1')\right] \nonumber \\[6pt]
& \phantom{=..} \times \left[ \cancel{A_{ss  \alpha'}^*(\omega_2')   \ha_{s \alpha'}^\dagger(\omega_2')} + A_{si   \alpha'}^*(\omega_2')  \ha_{i \alpha'}(\omega_{p \alpha'} - \omega_2') \right. \nonumber \\[6pt]
& \phantom{= \bra{0} [].} \left. + \cancel{ F_{s s \alpha'}^*(\omega_2')   \hf_{s \alpha'}^\dagger(\omega_2')} + F_{s i \alpha'}^*(\omega_2')  \hf_{i 1}(\omega_{p \alpha'} - \omega_2')\right] \nonumber \\[6pt]
& \phantom{=..} \times \left[ \cancel{A_{ss  \alpha}(\omega_2)   \ha_{s \alpha}(\omega_2)} + A_{si   \alpha}(\omega_2)  \ha^\dagger_{i \alpha}(\omega_{p \alpha} - \omega_2) \right. \nonumber \\[6pt]
& \phantom{= \bra{0} [].} \left. + \cancel{ F_{s s \alpha}(\omega_2)   \hf_{s \alpha}(\omega_2) } + F_{s i \alpha}(\omega_2)  \hf^\dagger_{i \alpha}(\omega_{p \alpha} - \omega_2)\right] \nonumber \\[6pt]
& \phantom{=..} \times \left[ A_{si   \beta}(\omega_1)  \ha^\dagger_{i \beta}(\omega_{p \beta} - \omega_1)  + F_{s i \beta}(\omega_1)  \hf^\dagger_{i \beta}(\omega_{p \beta} - \omega_1)\right] \ket{0} .
\end{align}
After removing the deleted terms, we obtain
\begin{align}\label{d190}
\langle \hbd_{s\beta'}(\omega_1') \hbd_{s\alpha'}(\omega_2') \hb_{s\alpha}(\omega_2) \hb_{s\beta}(\omega_1) \rangle & = \bra{0} \left[ A_{si   \beta'}^*(\omega_1')  \ha_{i \beta'}(\omega_{p \beta'} - \omega_1') + F_{s i \beta'}^*(\omega_1')  \hf_{i \beta'}(\omega_{p \beta'} - \omega_1')\right] \nonumber \\[6pt]
& \phantom{=..} \times \left[  A_{si   \alpha'}^*(\omega_2')  \ha_{i \alpha'}(\omega_{p \alpha'} - \omega_2') + F_{s i \alpha'}^*(\omega_2')  \hf_{i \alpha'}(\omega_{p \alpha'} - \omega_2')\right] \nonumber \\[6pt]
& \phantom{=..} \times \left[  A_{si   \alpha}(\omega_2)  \ha^\dagger_{i \alpha}(\omega_{p \alpha} - \omega_2) + F_{s i \alpha}(\omega_2)  \hf^\dagger_{i \alpha}(\omega_{p \alpha} - \omega_2)\right] \nonumber \\[6pt]
& \phantom{=..} \times \left[ A_{si   \beta}(\omega_1)  \ha^\dagger_{i \beta}(\omega_{p \beta} - \omega_1)  + F_{s i \beta}(\omega_1)  \hf^\dagger_{i \beta}(\omega_{p \beta} - \omega_1)\right] \ket{0} .
\end{align}
For practical calculations, it is convenient to rewrite in symbolic manner \eqref{d190} as,
\begin{align}\label{d200}
\langle \hbd_{s\beta'}(\omega_1') \hbd_{s\alpha'}(\omega_2') \hb_{s\alpha}(\omega_2) \hb_{s\beta}(\omega_1) \rangle & =  \bigl( \bra{1;\omega_1',\omega_2'} +  \bra{2;\omega_1',\omega_2'} + \bra{3;\omega_2',\omega_1'}+  \bra{4;\omega_1',\omega_2'} \bigr) \nonumber \\[6pt]
& \phantom{=} \times \bigl( \ket{1;\omega_1,\omega_2} +  \ket{2;\omega_1,\omega_2} + \ket{3;\omega_2,\omega_1}+  \ket{4;\omega_1,\omega_2} \bigr) ,
\end{align}
where we have defined,
\begin{equation}\label{d210}
\begin{split}
\bra{1;\omega_1',\omega_2'} & =  A_{si   \beta'}^*(\omega_1') A_{si   \alpha'}^*(\omega_2')  \bra{0}  \ha_{i \beta'}(\omega_{p \beta'} - \omega_1') \ha_{i \alpha'}(\omega_{p \alpha'} - \omega_2'),  \\[6pt]
\bra{2;\omega_1',\omega_2'} & =  A_{si   \beta'}^*(\omega_1') F_{s i \alpha'}^*(\omega_2')  \bra{0}  \ha_{i \beta'}(\omega_{p \beta'} - \omega_1') \hf_{i \alpha'}(\omega_{p \alpha'} - \omega_2'),  \\[6pt]
\bra{3;\omega_2',\omega_1'} & =  A_{si   \alpha'}^*(\omega_2') F_{s i \beta'}^*(\omega_1')  \bra{0}  \ha_{i \alpha'}(\omega_{p \alpha'} - \omega_2') \hf_{i \beta'}(\omega_{p \beta'} - \omega_1'),  \\[6pt]
\bra{4;\omega_1',\omega_2'} & =  F_{s i \beta'}^*(\omega_1') F_{s i \alpha'}^*(\omega_2')  \bra{0}  \hf_{i \beta'}(\omega_{p \beta'} - \omega_1') \hf_{i \alpha'}(\omega_{p \alpha'} - \omega_2'),
\end{split}
\end{equation}
and
\begin{equation}\label{d220}
\begin{split}
\ket{1;\omega_1,\omega_2} & =  A_{si   \alpha} (\omega_2) A_{si   \beta} (\omega_1)  \had_{i \alpha}(\omega_{p \alpha} - \omega_2) \had_{i \beta}(\omega_{p \beta} - \omega_1)\ket{0} ,  \\[6pt]
\ket{2;\omega_2,\omega_1} & =  A_{si   \beta} (\omega_1) F_{s i \alpha} (\omega_2)  \had_{i \beta}(\omega_{p \beta} - \omega_1) \hfd_{i \alpha}(\omega_{p \alpha} - \omega_2)\ket{0} ,  \\[6pt]
\ket{3;\omega_1,\omega_2} & =  A_{si   \alpha} (\omega_2) F_{s i \beta} (\omega_1)  \had_{i \alpha}(\omega_{p \alpha} - \omega_2) \hfd_{i \beta}(\omega_{p \beta} - \omega_1)\ket{0} ,  \\[6pt]
\ket{4;\omega_1,\omega_2} & =  F_{s i \alpha} (\omega_2) F_{s i \beta} (\omega_1)  \hfd_{i \alpha}(\omega_{p \alpha} - \omega_2) \hfd_{i \beta}(\omega_{p \beta} - \omega_1)\ket{0} .
\end{split}
\end{equation}
It should be clear now that the quantity $\langle \hbd_{s\beta'}(\omega_1') \hbd_{s\alpha'}(\omega_2') \hb_{s\alpha}(\omega_2) \hb_{s\beta}(\omega_1) \rangle$ contains only the following six terms:
\begin{align}\label{d230}
\langle \hbd_{s\beta'}(\omega_1') \hbd_{s\alpha'}(\omega_2') \hb_{s\alpha}(\omega_2) \hb_{s\beta}(\omega_1) \rangle & = \brak{1; \omega_1', \omega_2'}{1; \omega_1, \omega_2} +
\brak{2; \omega_1', \omega_2'}{2; \omega_2, \omega_1} \nonumber \\[6pt]
& \phantom{=.} + \brak{3; \omega_2', \omega_1'}{3; \omega_1, \omega_2}+\brak{4; \omega_1', \omega_2'}{4; \omega_1, \omega_2}, \nonumber \\[6pt]
& \phantom{=.} + \brak{2; \omega_1', \omega_2'}{3; \omega_1, \omega_2} + \brak{3; \omega_2', \omega_1'}{2; \omega_2, \omega_1}.
\end{align}
A straightforward calculation gives,
\begin{equation}\label{d240}
\begin{split}
\brak{1; \omega_1', \omega_2'}{1; \omega_1, \omega_2} & = A_{si   \beta'}^*(\omega_1') A_{si   \alpha'}^*(\omega_2') A_{si   \alpha} (\omega_2) A_{si   \beta} (\omega_1) \nonumber \\[6pt]
& \phantom{=} \times \underbrace{\bra{0}  \ha_{i \beta'}(\omega_{p \beta'} - \omega_1') \ha_{i \alpha'}(\omega_{p \alpha'} - \omega_2') \had_{i \alpha}(\omega_{p \alpha} - \omega_2) \had_{i \beta}(\omega_{p \beta} - \omega_1)\ket{0}}_{= \, \delta_{\alpha \alpha'}\delta_{\beta \beta'} \delta(\omega_1' - \omega_1) \delta(\omega_2'-\omega_2)   +  \delta_{\alpha \beta'}\delta_{\beta \alpha'} \delta(\omega_1' - \omega_2) \delta(\omega_2'-\omega_1 )} , \\[6pt]
\brak{2; \omega_1', \omega_2'}{2; \omega_2, \omega_1} & =  A_{si   \beta'}^*(\omega_1') F_{s i \alpha'}^*(\omega_2')  A_{si   \beta} (\omega_1) F_{s i \alpha} (\omega_2)  \nonumber \\[6pt]
& \phantom{=} \times \underbrace{\bra{0}  \ha_{i \beta'}(\omega_{p \beta'} - \omega_1') \hf_{i \alpha'}(\omega_{p \alpha'} - \omega_2') \had_{i \beta}(\omega_{p \beta} - \omega_1) \hfd_{i \alpha}(\omega_{p \alpha} - \omega_2)\ket{0}}_{= \, \delta_{\alpha \alpha'}\delta_{\beta \beta'} \delta(\omega_1' - \omega_1) \delta(\omega_2'-\omega_2) },  \\[6pt]
 \brak{3; \omega_2', \omega_1'}{3; \omega_1, \omega_2} & =  A_{si   \alpha'}^*(\omega_2') F_{s i \beta'}^*(\omega_1') A_{si   \alpha} (\omega_2) F_{s i \beta} (\omega_1) \nonumber \\[6pt]
& \phantom{=} \times  \underbrace{ \bra{0}  \ha_{i \alpha'}(\omega_{p \alpha'} - \omega_2') \hf_{i \beta'}(\omega_{p \beta'} - \omega_1') \had_{i \alpha}(\omega_{p \alpha} - \omega_2) \hfd_{i \beta}(\omega_{p \beta} - \omega_1)\ket{0}}_{= \, \delta_{\alpha \alpha'}\delta_{\beta \beta'} \delta(\omega_1' - \omega_1) \delta(\omega_2'-\omega_2) } ,  \\[6pt]
\brak{4; \omega_1', \omega_2'}{4; \omega_1, \omega_2} & =  F_{s i \beta'}^*(\omega_1') F_{s i \alpha'}^*(\omega_2')F_{s i \alpha} (\omega_2) F_{s i \beta} (\omega_1) \nonumber \\[6pt]
& \phantom{=} \times \underbrace{\bra{0}  \hf_{i \beta'}(\omega_{p \beta'} - \omega_1') \hf_{i \alpha'}(\omega_{p \alpha'} - \omega_2') \hfd_{i \alpha}(\omega_{p \alpha} - \omega_2) \hfd_{i \beta}(\omega_{p \beta} - \omega_1)\ket{0}}_{= \, \delta_{\alpha \alpha'}\delta_{\beta \beta'} \delta(\omega_1' - \omega_1) \delta(\omega_2'-\omega_2)   +  \delta_{\alpha \beta'}\delta_{\beta \alpha'} \delta(\omega_1' - \omega_2) \delta(\omega_2'-\omega_1) } , \\[6pt]
\brak{2; \omega_1', \omega_2'}{3; \omega_1, \omega_2}  & =   A_{si   \beta'}^*(\omega_1') F_{s i \alpha'}^*(\omega_2') A_{si   \alpha} (\omega_2) F_{s i \beta} (\omega_1) \nonumber \\[6pt]
& \phantom{=} \times \underbrace{\bra{0}  \ha_{i \beta'}(\omega_{p \beta'} - \omega_1') \hf_{i \alpha'}(\omega_{p \alpha'} - \omega_2') \had_{i \alpha}(\omega_{p \alpha} - \omega_2) \hfd_{i \beta}(\omega_{p \beta} - \omega_1)\ket{0} }_{= \, \delta_{\alpha \beta'}\delta_{\beta \alpha'} \delta(\omega_1' - \omega_2) \delta(\omega_2'-\omega_1)},  \\[6pt]
\brak{3; \omega_2', \omega_1'}{2; \omega_2, \omega_1} & =  A_{si   \alpha'}^*(\omega_2') F_{s i \beta'}^*(\omega_1') A_{si   \beta} (\omega_1) F_{s i \alpha} (\omega_2) \nonumber \\[6pt]
& \phantom{=} \times \underbrace{\bra{0}  \ha_{i \alpha'}(\omega_{p \alpha'} - \omega_2') \hf_{i \beta'}(\omega_{p \beta'} - \omega_1') \had_{i \beta}(\omega_{p \beta} - \omega_1) \hfd_{i \alpha}(\omega_{p \alpha} - \omega_2)\ket{0} }_{= \, \delta_{\alpha \beta'}\delta_{\beta \alpha'} \delta(\omega_1' - \omega_2) \delta(\omega_2'-\omega_1)} .
\end{split}
\end{equation}
Using the properties of the Dirac distribution we can rewrite the equations above as,
\begin{equation}\label{d250}
\begin{split}
\brak{1; \omega_1', \omega_2'}{1; \omega_1, \omega_2} & =  \abs{A_{si   \alpha} (\omega_2) A_{si   \beta} (\omega_1) }^2 \nonumber \\[6pt]
& \phantom{=} \times \left[ \delta_{\alpha \alpha'}\delta_{\beta \beta'} \delta(\omega_1' - \omega_1) \delta(\omega_2'-\omega_2)   +  \textcolor{blue}{\delta_{\alpha \beta'}\delta_{\beta \alpha'} \delta(\omega_1' - \omega_2) \delta(\omega_2'-\omega_1 )} \right] , \\[6pt]
\brak{2; \omega_1', \omega_2'}{2; \omega_2, \omega_1} & =  \abs{  A_{si   \beta} (\omega_1) F_{s i \alpha} (\omega_2) }^2 \delta_{\alpha \alpha'}\delta_{\beta \beta'} \delta(\omega_1' - \omega_1) \delta(\omega_2'-\omega_2) \nonumber \\[6pt]
 \brak{3; \omega_2', \omega_1'}{3; \omega_1, \omega_2} & =  \abs{ A_{si   \alpha} (\omega_2) F_{s i \beta} (\omega_1)}^2 \delta_{\alpha \alpha'}\delta_{\beta \beta'} \delta(\omega_1' - \omega_1) \delta(\omega_2'-\omega_2)  \nonumber \\[6pt]
\brak{4; \omega_1', \omega_2'}{4; \omega_1, \omega_2} & =  \abs{ F_{s i \alpha} (\omega_2) F_{s i \beta} (\omega_1)}^2 \nonumber \\[6pt]
& \phantom{=} \times \left[\delta_{\alpha \alpha'}\delta_{\beta \beta'} \delta(\omega_1' - \omega_1) \delta(\omega_2'-\omega_2)   +  \textcolor{blue}{\delta_{\alpha \beta'}\delta_{\beta \alpha'} \delta(\omega_1' - \omega_2) \delta(\omega_2'-\omega_1) }\right] , \\[6pt]
\brak{2; \omega_1', \omega_2'}{3; \omega_1, \omega_2}  & = \abs{ A_{si   \alpha} (\omega_2) F_{s i \beta} (\omega_1)}^2 \textcolor{blue}{\delta_{\alpha \beta'}\delta_{\beta \alpha'} \delta(\omega_1' - \omega_2) \delta(\omega_2'-\omega_1)}  \nonumber \\[6pt]
\brak{3; \omega_2', \omega_1'}{2; \omega_2, \omega_1} & = \abs{A_{si   \beta} (\omega_1) F_{s i \alpha} (\omega_2)}^2 \textcolor{blue}{\delta_{\alpha \beta'}\delta_{\beta \alpha'} \delta(\omega_1' - \omega_2) \delta(\omega_2'-\omega_1)},
\end{split}
\end{equation} 
where the terms highlighted in \textcolor{blue}{blue} are the ones responsible for the time dependence of the second-order correlation function.

Substituting this results in \eqref{d170}, we find
\begin{align}\label{d260}
\langle  \hdd_{s1}(\omega_1')  \hdd_{s1}(\omega_2') \hd_{s1}(\omega_2)  \hd_{s1}(\omega_1) \rangle & = \delta(\omega_1' - \omega_1) \delta(\omega_2'-\omega_2) 
   \nonumber \\[6pt]
& \phantom{=.} \times \sum_{\alpha,\beta=1}^2 \abs{ G_{\alpha \beta}}^2 \Bigl[\abs{A_{si   \alpha} (\omega_2) A_{si   \beta} (\omega_1) }^2+ \abs{  A_{si   \beta} (\omega_1) F_{s i \alpha} (\omega_2) }^2 \nonumber \\[6pt]
& \phantom{=.} \phantom{\times .} +   \abs{ A_{si   \alpha} (\omega_2) F_{s i \beta} (\omega_1)}^2 +  \abs{ F_{s i \alpha} (\omega_2) F_{s i \beta} (\omega_1)}^2 \Bigr]    \nonumber \\[6pt]
& \phantom{=.} +  \textcolor{blue}{\delta(\omega_1' - \omega_2) \delta(\omega_2'-\omega_1)}   \nonumber \\[6pt]
& \phantom{=.} \times \sum_{\alpha,\beta=1}^2 G_{\beta \alpha}^* G_{\alpha \beta } \Bigl[\abs{A_{si   \alpha} (\omega_2) A_{si   \beta} (\omega_1) }^2 + \abs{A_{si   \beta} (\omega_1) F_{s i \alpha} (\omega_2)}^2  \Bigr] \nonumber \\[6pt]
& \phantom{=.} \phantom{\times .} +   \abs{ A_{si   \alpha} (\omega_2) F_{s i \beta} (\omega_1)}^2 +  \abs{ F_{s i \alpha} (\omega_2) F_{s i \beta} (\omega_1)}^2  \Bigr]    
\end{align}
From \eqref{d140} we can see that 
\begin{align}\label{d270}
e^{i \tau_1(\omega_1 - \omega_1')  + i \tau_2 (\omega_2 - \omega_2'  )} \times \left\{
\begin{array}{ll}
\delta(\omega_1' - \omega_1) \delta(\omega_2'-\omega_2) & = 1,  \\[6pt]
\textcolor{blue}{\delta(\omega_1' - \omega_2) \delta(\omega_2'-\omega_1)} & =  e^{-i \tau \omega_1 }  e^{ i \tau \omega_2} ,
\end{array}
\right.
\end{align}
where $\tau = \tau_2 - \tau_1$. Now, substituting (\ref{d260}-\ref{d270}) into \eqref{d140}, and remembering the definitions \eqref{d100}, we can eventually write
\begin{align}\label{d280}
\langle \hdd_{s1}(\tau_1) \hdd_{s1}(\tau_2) \hd_{s1}(\tau_2) \hd_{s1}(\tau_1) \rangle & =  \sum_{\alpha,\beta=1}^2 \abs{ G_{\alpha \beta}}^2 \Bigl[  \varphi_{A_\alpha}(0) \varphi_{A_\beta}(0) + \varphi_{A_\beta}(0)\varphi_{F_\alpha}(0) \nonumber \\[6pt]
& \phantom{=.} +  \varphi_{A_\alpha}(0)\varphi_{F_\beta}(0) + \varphi_{F_\alpha}(0) \varphi_{F_\beta}(0) \Bigr]    \nonumber \\[6pt]
& \phantom{=.} + \sum_{\alpha,\beta=1}^2  G_{\beta \alpha}^* G_{\alpha \beta } \Bigl[ \varphi_{A_\alpha}^*(\tau) \varphi_{A_\beta}(\tau)  + \varphi_{A_\beta}(\tau)\varphi_{F_\alpha}^*(\tau)   \Bigr] \nonumber \\[6pt]
& \phantom{=.} \phantom{\times .} +    \varphi_{A_\alpha}^* (\tau)\varphi_{F_\beta}(\tau) + \varphi_{F_\alpha}^* (\tau) \varphi_{F_\beta}(\tau) \Bigr]   . 
\end{align}
After rearranging some terms in the equation above and using the definition \eqref{d110}, we find
\begin{align}\label{d290}
\langle \hdd_{s1}(\tau_1) \hdd_{s1}(\tau_2) \hd_{s1}(\tau_2) \hd_{s1}(\tau_1) \rangle & =  \sum_{\alpha,\beta=1}^2 \abs{ G_{\alpha \beta}}^2 \bigl[  \varphi_{A_\alpha}(0) + \varphi_{F_\alpha}(0) \bigl] \bigl[  \varphi_{F_\beta}(0) + \varphi_{F_\alpha}(0) \bigl]    \nonumber \\[6pt]
& \phantom{=.} + \sum_{\alpha,\beta=1}^2  G_{\beta \alpha}^* G_{\alpha \beta } \bigl[  \varphi_{A_\alpha}^*(\tau) + \varphi_{F_\alpha}^*(\tau) \bigl] \bigl[  \varphi_{A_\beta}(\tau) + \varphi_{F_\beta}(\tau)\bigl]  \nonumber \\[6pt]
& = \sum_{\alpha,\beta=1}^2 \abs{ G_{\alpha \beta}}^2 \phi_\alpha(0) \phi_\beta(0)  + \sum_{\alpha,\beta=1}^2  G_{\beta \alpha}^* G_{\alpha \beta } \phi_\alpha^*(\tau) \phi_\beta(\tau) . 
\end{align}
Now, we are ready to use the definition \eqref{d160} to rewrite  $\langle \hdd_{s1}(\tau_1) \hdd_{s1}(\tau_2) \hd_{s1}(\tau_2) \hd_{s1}(\tau_1) \rangle$ as,
\begin{align}\label{d300}
\langle \hdd_{s1}(\tau_1) \hdd_{s1}(\tau_2) \hd_{s1}(\tau_2) \hd_{s1}(\tau_1) \rangle & =  \Bigl[\abs{\mT}^2 \phi_1(0) + \abs{\maR}^2 \phi_2(0) \Bigr] \Bigl[\abs{\maR}^2 \phi_1(0) + \abs{\mT}^2 \phi_2(0) \Bigr]  \nonumber \\[6pt]
& \phantom{=.} + \abs{\maR \mT}^2 \abs{  \phi_{1}(\tau) - \phi_{2}(\tau)}^2 ,
\end{align}
where we have used \eqref{d10} to derive $\left( \maR \mT^* \right)^2 = - \abs{ \maR \mT }^2$.
Finally, substituting \eqref{d130} and\eqref{d300} into \eqref{d30}, we obtain
\begin{align}\label{d30bis}
g^{(2)} \left( \tau \right) = 1 + \abs{\maR \mT}^2 \dfrac{ \abs{  \phi_{1}(\tau) - \phi_{2}(\tau)}^2}{\Bigl[\abs{\mT}^2 \phi_1(0) + \abs{\maR}^2 \phi_2(0) \Bigr] \Bigl[\abs{\maR}^2 \phi_1(0) + \abs{\mT}^2 \phi_2(0) \Bigr]},
\end{align}
which is the sought result. In explicit form, we have
\begin{align}\label{g2}
g^{(2)} \left( \tau \right)  = 1 + 2\abs{\maR \mT}^2 \frac{I^2(\tau)}{I^2(0)} \bigl[ 1 - \cos \left(\delta \omega \, \tau \right)\bigr],
\end{align}
where we have defined the small frequency detuning $\delta \omega$ between the two thermal sources, as
\begin{align}\label{deltaomega}
\delta \omega \deff \omega_{s2} - \omega_{s1}.
\end{align}

\section{Frequency tuning}

\begin{figure} [hbpt!]
    \centering
    \includegraphics[width=.8\linewidth]{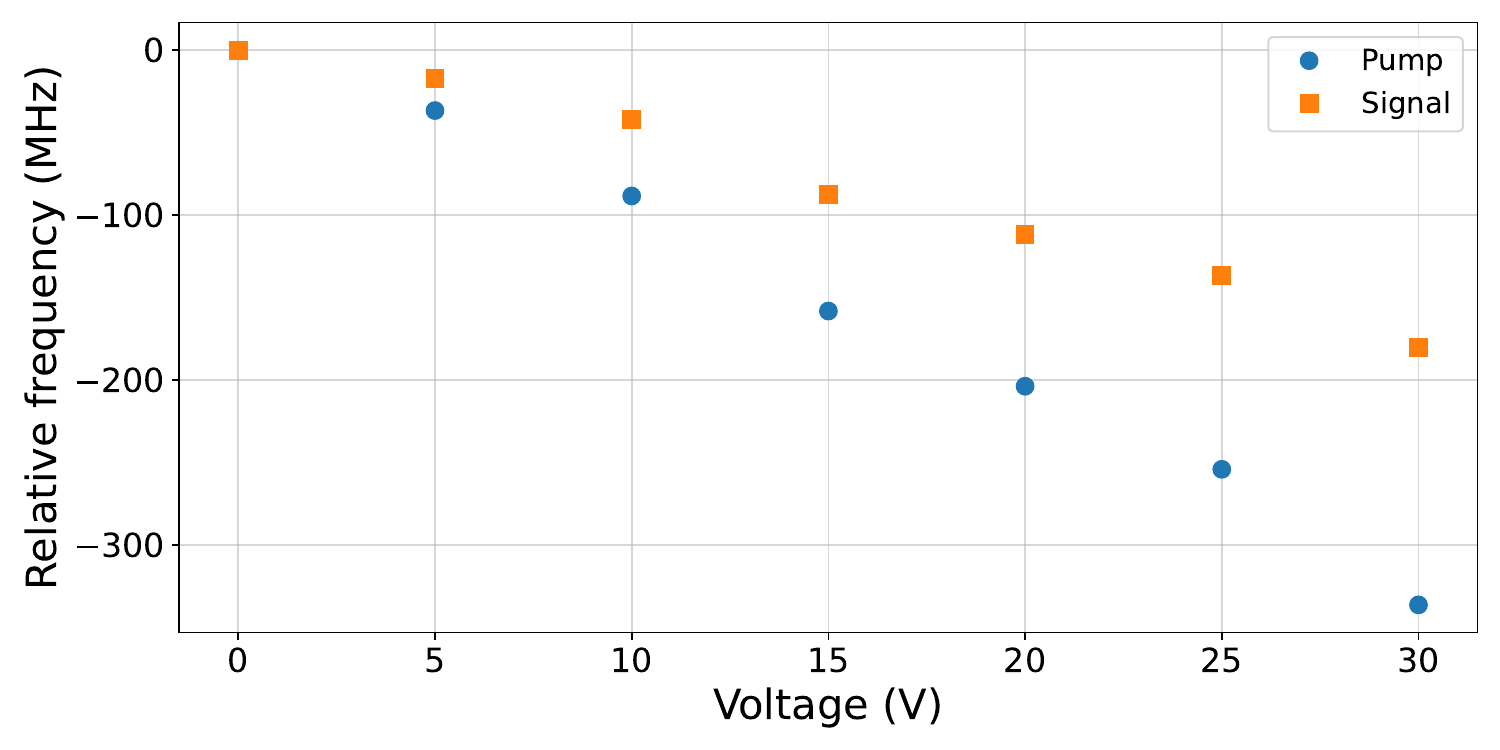}
    \caption{Frequency shift of the pump and signal beams with the voltage applied to the nanopositioner relative to the initial frequency measured for an applied voltage of 0\,V.}
    \label{fig:WGMR_tuning}
\end{figure}

Since WGMRs are monolithic, triply resonant cavity-enhanced SPDC sources, changing the temperature of the WGMR results in a discrete frequency shift of the generated beams. For the WGMR used in this experiment, temperature tuning results in a discrete frequency shift of about \qty{28}{GHz}. To access a finer frequency tuning resolution, we implement a technique that allows for continuous frequency tuning. An aluminum pin attached to a piezo actuator is mounted on top of the second WGMR. By applying voltage to the piezo actuator, the pin moves toward the WGMR, mechanically deforming it and slightly increasing its size. As a result, we induce a frequency redshift using this technique~\cite{pollinger2009ultrahigh}. To measure the influence of the deformation to the phase matching condition, we conduct the followings experiment. We first stabilize the pump frequency at the maximum count rate of the downconversion process. An auxiliary laser at the signal wavelength is coupled into the WGMR. By finding the resonance frequency of the laser, we can determine the signal frequency. We then apply a voltage to the nanopositioner, causing the photon count rate to drop because the phase-matching condition is altered. Finally, we tune the pump frequency until a maximum count rate is achieved again and record both the pump frequency and the corresponding signal frequency. The relationship between the applied voltage and the frequency shift of the pump laser and the signal beam are shown in Fig.~\ref{fig:WGMR_tuning}. The results show that shifting the pump frequency by \qty{1}{MHz} results in a shift of approximately \qty{0.55}{MHz} in the signal. This allows us to determine the frequency shift of the signal simply by observing the shift in the pump.

\section{Signal-idler second-order correlation function}

The parameters, \(\Gamma_s\), \(\Gamma_i\), and \(\kappa_1\) can be extracted from the cross-correlation functions measured between the signal and idler. Follow the calculation provided above, the cross-correlation functions can be expressed as

\begin{align}
g_{is}^{(2)}(\tau) &= \frac{\langle \hat{b}_i^\dagger(\tau_1) \hat{b}_s^\dagger(\tau_2) \hat{b}_s(\tau_2) \hat{b}_i(\tau_1) \rangle}{\langle \hat{b}_i^\dagger(\tau_1) \hat{b}_i(\tau_1) \rangle \langle \hat{b}_s^\dagger(\tau_2) \hat{b}_s(\tau_2) \rangle} \\[1.5ex]
&= 1 + \frac{\Gamma_i \Gamma_s}{\kappa_1^2} \left[ \frac{I_1(\tau)}{I_0} - \Gamma_s \frac{I_2(\tau)}{I_0} \right]^2,
\end{align}
where we have defined
\begin{align*}
I_0 &= \frac{1}{2\pi} \int_{\mathbb{R}} \frac{1}{|D(x)|^2} \,\mathrm{d}x \\
&= \frac{4}{\Gamma_i \Gamma_s - 4\kappa_1^2} \frac{1}{ (a_+ + a_-)},
\end{align*}
where
\[
D(x) = (\Gamma_i/2 - ix)(\Gamma_s/2 - ix) - \kappa_1^2.
\]
the remaining two integrals are
\begin{align*}
I_1(\tau) &= \frac{1}{2\pi} \int_{\mathbb{R}} \frac{1}{D(x)} e^{ix\tau} \,\mathrm{d}x \\
&= -\operatorname{H}(-\tau) \frac{4 \, e^{(\Gamma_i + \Gamma_s)/4}}{\sqrt{(\Gamma_i - \Gamma_s)^2 + 16\kappa_1^2}} \sinh \left[ \frac{\sqrt{(\Gamma_i - \Gamma_s)^2 + 16\kappa_1^2}}{4} \tau \right],
\end{align*}
where $\operatorname{H}(t)$ is the Heaviside step function, and
\begin{align*}
I_2(\tau) &= \frac{1}{2\pi} \int_{\mathbb{R}} \frac{\Gamma_i/2 + ix}{|D(x)|^2} e^{ix\tau} \,\mathrm{d}x \\
&= -\frac{2\Gamma_i}{\sqrt{a_+^2-a_-^2}} \left( \frac{e^{-a_+ |\tau|/2}}{a_+} - \frac{e^{-a_- |\tau|/2}}{a_-} \right) \\
&\quad + \frac{2\operatorname{sign}(\tau)}{\sqrt{a_+^2-a_-^2}} \left( e^{-a_+ |\tau|/2} - e^{-a_- |\tau|/2} \right).
\end{align*}

\begin{figure}[ht!]
    \centering
    \includegraphics[width=.8\linewidth]{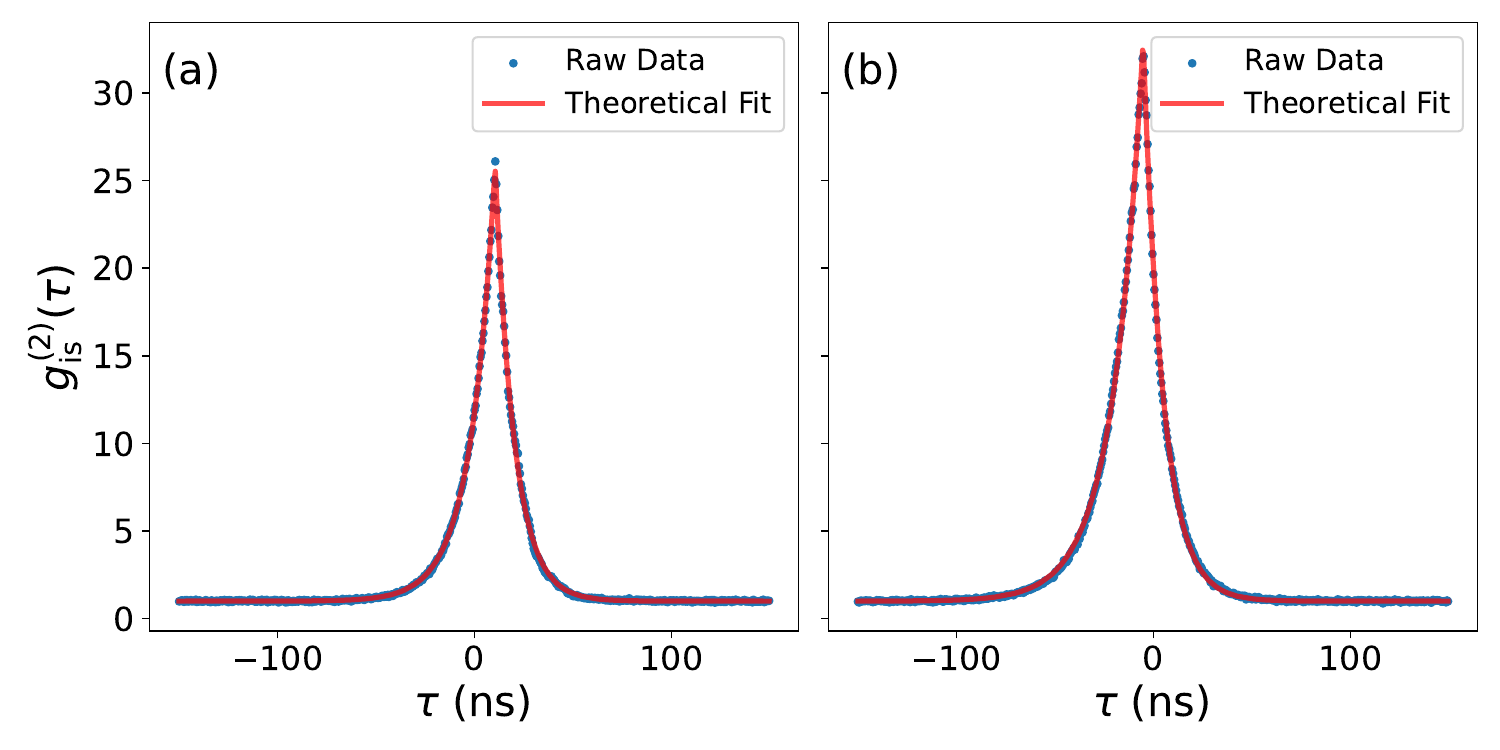}
    \caption{Cross-correlation functions of the two WGMRs. (a) Data measured from WGMR1. (b) Data measured from WGMR2. The blue dots represent the experimental results, while the red line represents the theoretical fit.}
    \label{fig:g2_is}
\end{figure}

Fig.~\ref{fig:g2_is} shows the cross-correlation functions of the two WGMRs measured at zero detuning. Using the model provided above, we find that $\Gamma_{\mathrm{s}}\approx0.08$, $\Gamma_{\mathrm{i}}\approx0.10$, and $\kappa_1\approx7.3\times10^{-5}$ for WGMR1, and $\Gamma_{\mathrm{s}}\approx0.07$, $\Gamma_{\mathrm{i}}\approx0.10$, and $\kappa_1\approx4.6\times10^{-5}$ for WGMR2. Note that these values are extracted from the same measurements shown in Fig.~5 of the main manuscript. This implies that the actual $g^{(2)}_{\mathrm{is}}(0)$ of the two WGMRs is higher than what is displayed in the figure. A higher $g^{(2)}_{\mathrm{is}}(0)$ will, in turn, result in a lower $\kappa_1$. Given that the $4\kappa_1^2$ term obtained here is already five orders of magnitude less than the product $\Gamma_{\mathrm{s}}\Gamma_{\mathrm{i}}$, we can safely use the approximate result of Eq.~(18) from the main manuscript. We extracted the cross-correlation functions from all measurements presented in Fig.~5 of the main manuscript to ensure that the parameters used for our theoretical model are correct for each detuning case.

\bibliography{HOM_bib}